\documentstyle[epsfig,12pt,here]{article}
\newcommand{\mycomm}[1]{\hfill\break
$\phantom{a}$\kern-3.5em{\tt===$>$ \bf #1}\hfill\break}
\newcommand{\mycommA}[1]{\hfill\break
$\phantom{a}$\kern-3.5em{\tt***$>$ \bf #1}\hfill\break}

\begin{document}
\catcode`\@=11 
\def\lsim{\mathrel{\mathpalette\@versim<}}
\def\gsim{\mathrel{\mathpalette\@versim>}}
\def\@versim#1#2{\vcenter{\offinterlineskip
        \ialign{$\m@th#1\hfil##\hfil$\crcr#2\crcr\sim\crcr } }}
\catcode`\@=12 
\def\beq{\begin{equation}}
\def\eeq{\end{equation}}
\def\MSbar {\hbox{$\overline{\hbox{\tiny MS}}\,$}}
\def\eff{\hbox{\tiny eff}}
\def\FP{\hbox{\tiny FP}}
\def\IR{\hbox{\tiny IR}}
\def\UV{\hbox{\tiny UV}}
\def\APT{\hbox{\tiny APT}}
\def\DRED{\hbox{\tiny DRED}}
\def\PA{\hbox{\tiny PA}}
\def\mysim{\kern -.1667em\lower0.8ex\hbox{$\tilde{\phantom{a}}$}}
\vskip 20pt

\begin{titlepage}
\begin{flushright}
{\footnotesize
S645.0998\\}
\end{flushright}
\begin{centering}
\vspace{0.8in}
{\large{\bf
The conformal window in QCD and supersymmetric QCD 
}}
\vskip 30pt
\vskip 30pt
{\bf Einan Gardi} \,\,\, and \,\,\,{\bf Georges Grunberg}

\vspace{.4in}

 Centre de Physique Th\'eorique de l'Ecole Polytechnique\footnote{CNRS
UMR C7644}
\\ 91128 Palaiseau Cedex, France
\\email: gardi@cpht.polytechnique.fr,\, grunberg@cpht.polytechnique.fr

\vspace{0.9in}

{\bf Abstract} \\
\vspace{0.35cm}
{\small
In both QCD and supersymmetric QCD (SQCD) with $N_f$ flavors there are
conformal windows where the theory is asymptotically free in the ultraviolet
while the infrared physics is governed by a non-trivial fixed-point.
In SQCD, the lower $N_f$ boundary of the conformal window, below which
the theory is confining is well understood thanks to duality. 
In QCD there is just a sufficient condition for confinement based on 
superconvergence.
Studying the Banks-Zaks expansion and analyzing the conditions for
the perturbative coupling to have a {\em causal analyticity structure},
it is shown that the infrared fixed-point in QCD is perturbative
in the entire conformal window. This finding suggests 
that there can be no analog of duality in QCD.  
On the other hand in SQCD the infrared region is found to be strongly 
coupled in the lower part of the conformal window, in agreement with duality.
Nevertheless, we show that it is possible to interpolate
between the Banks-Zaks expansions in the electric and magnetic
theories, for quantities that can be calculated perturbatively in both.
This interpolation is explicitly demonstrated for
the critical exponent that controls the rate at which a generic
physical quantity approaches the fixed-point.} 
\end{centering}

\end{titlepage}
\vfill\eject

\section{Introduction}

In multi-flavor QCD, there is a conformal window \cite{BZ}, namely a region of
$N_f$ values for which the theory is asymptotically free at short distances 
while the long distance physics is governed by a non-trivial
fixed-point. This is a non-Abelian Coulomb phase in which quarks and
gluons are not confined.

The upper boundary of the conformal window is determined according to
the sign of the $\beta$ function 
\beq
\beta(x)\,\equiv\, \frac{dx}{d\ln(\mu^2)}\,=\,-
\left(\beta_0x^2+\beta_1x^3+\cdots\right), 
\eeq
at small coupling $x\equiv\alpha_s/\pi$. 
When the first coefficient of the perturbative $\beta$ function \cite{oneloop},
\beq
\beta _0=\frac 14\left( \frac{11}{3}N_c-\frac 23N_f\right),
\label{beta0}
\eeq
changes its sign the theory changes its nature from the asymptotically
free phase
\hbox{$r\equiv N_f/N_c<11/2$}, to the infrared free phase \hbox{$r>11/2$}.   
The transition point in $N_c=3$ QCD is at $N_f=16.5$. For
$r<11/2$ the $\beta$ function is negative for 
a vanishingly small coupling ($\beta_0>0$), but due to the second term 
which has an opposite sign ($\beta_1<0$), $\beta(x)$ reaches a
non-trivial zero at some \hbox{$x_{\FP}\simeq-\beta_0/\beta_1>0$} 
\cite{twoloops}. 
$x_{\FP}$ approaches zero as $r$ approaches $11/2$, and then quarks
and gluons are weakly coupled at all scales. Finally, the smallness of
$x_{\FP}$ justifies the use of the 2-loop $\beta$ function.

On the other hand, the lower boundary of the conformal window, below
which confinement sets in, is much harder to tackle. One approach to 
confinement, the so-called metric confinement \cite{Oehme_metric_confinement}, 
defines confinement as a phase in which transverse gauge field
excitations are excluded from the space of physical states which is
defined through the BRST algebra.
It was shown in \cite{Oehme_metric_confinement} that as long as a
certain condition is obeyed by the gauge field propagator, metric
confinement is implied. This condition is most conveniently expressed
in the Landau gauge, namely that the absorptive part of the gluon
propagator $D(Q^2,\mu^2,g)$ (in this gauge) is superconvergent,
\beq
\int_{0^-}^{\infty}dk^2\,\rho(k^2,\mu^2,g)\,=\,0,
\label{superconvergence_relation}
\eeq
where $\mu^2$ is the renormalization scale and
$\rho(k^2,\mu^2,g)=(1/\pi)\,{\rm Im}\left\{D(-k^2,\mu^2,g)\right\}$.  

Assuming analyticity of the gluon propagator $D(Q^2,\mu^2,g)$, the
superconvergence relation (\ref{superconvergence_relation}) was shown
to be a direct consequence of renormalization group invariance,
provided $D(Q^2,\mu^2,g)$ vanishes faster than $1/Q^2$ at 
large $Q^2$.  Due to asymptotic freedom, the last condition depends
only on the sign of the 1-loop anomalous dimension of the propagator
(in the Landau gauge), given by
\beq
\gamma_{00}= -\frac 14\left( \frac{13}{6}N_c-\frac 23N_f\right).
\label{gamma00}
\eeq
If $\gamma_{00}$ is negative the superconvergence relation
(\ref{superconvergence_relation}) holds.

Thus, in this approach, a {\em sufficient} condition for confinement 
is that
$\gamma_{00}<0$ and therefore the lower boundary of the conformal
window cannot be lower than $r=13/4$. If superconvergence is also a
necessary condition (this has not been shown) then the phase
transition should be at $r=13/4$.  
In $N_c=3$ QCD this corresponds to $N_f=9.75$, i.e. between 9 and 10 flavors. 

A natural question to ask is whether the superconvergence condition
necessarily implies that also quarks are confined. To the knowledge of
the authors no complete answer has yet been given to this question, although   
it has been shown that the superconvergence criterion is
consistent with a potential that is approximately linear in 
some intermediate scales \cite{Oehme_potential_confinement}. 

There are several other approaches to study the phase structure of multi-flavor
QCD, such as the instanton liquid model \cite{Shuryak}, 
the gap equation \cite{Appelquist,Miransky} and computer simulations on the 
lattice \cite{lattice_old,lattice_Japan}. 
The new lattice results \cite{lattice_Japan} are inconsistent with old ones
\cite{lattice_old} and with the superconvergence criterion for 
confinement: they 
indicate that the phase where no confinement nor chiral symmetry
breakdown occurs, stretches down to $N_f=7$, and thus only for $6$ flavors
and below
QCD appears as a confining theory with chiral symmetry breaking, as we know it
in the real world. In spite of these contradicting evidence, we assume
in this paper that the bottom of the conformal window is as implied by
superconvergence.

Recently the presence of a fixed-point in QCD was studied as a
function $N_f$ by considering the perturbative $\beta$ function
\cite{FP,LamW}. Three approaches where considered: a direct
investigation\footnote{The relevant refs. appear in \cite{FP}.}
of the equation $\beta(x)=0$ in physical renormalization
schemes \cite{ECH}, the Banks-Zaks expansion \cite{BZ,BZ_grunberg,CaSt}  
and the analyticity structure of the coupling constant.

The direct investigation of zeros in the QCD $\beta$ function in physical
schemes \cite{FP} shows that at 3-loop, a fixed-point can appear for
most effective charges above $N_f\simeq 5$. However, presence of a fixed-point 
{\em at the lower end} is very sensitive to higher-loop corrections, and
thus cannot be trusted. 

The Banks-Zaks expansion is an expansion in the number of flavors
down from the point where $\beta_0$ changes its
sign. The expansion parameter is proportional to $\beta_0$, or in $N_c=3$
QCD to $(16.5-N_f)$. 
It was found in \cite{FP} that Banks-Zaks series for different QCD 
observables behave differently: in some cases the coefficients are
small and the expansion is reliable and in other cases it seems to 
breakdown at order $(16.5-N_f)^3$ already around 10 or 12 flavors (see
fig.~6 in \cite{FP}). Here we further interpret these results. 

In real-world QCD there are Landau singularities in the perturbative
coupling, which signal the inapplicability of perturbation theory to
describe the infrared physics. These non-physical singularities 
are usually assumed to be compensated by 
non-perturbative  power-like terms in any physical quantity. Thus
causality is realized only at the non-perturbative  level.
The existence of a perturbative fixed-point opens 
up the possibility that the perturbative
coupling will have no Landau singularities \cite{FP,LamW}. 
We will assume that {\em within the conformal window} non-perturbative 
effects are not important as long as they are not implied by
perturbation theory, that is as long as the perturbative coupling is
causal and small.
Note that the causality requirement is stronger than the requirement
to have no space-like Landau singularity.

The simplest example where causality of the coupling can be achieved at
the perturbative level, without additional power corrections,
is the 2-loop coupling.
In \cite{LamW} an exact explicit formula for the 2-loop coupling as
a function of the scale was introduced, which enabled a complete
understanding of the singularity structure of the coupling in this
approximation. It turns
out that the condition for causality of the 2-loop coupling is 
$c\equiv\beta_1/\beta_0<-\beta_0$. This condition translates in
$N_c=3$ QCD to $N_f\geq 10$. 

As stated above the lower boundary of the conformal window implied by
superconvergence is also between $9$ and $10$ flavors 
\cite{Oehme_metric_confinement}. Thus basing on 
the superconvergence criterion, we find that the 2-loop
perturbative coupling is causal in the entire conformal window.
This suggests that the perturbative analysis in the infrared
is reliable down to the bottom of the window.
On the other hand, perturbation theory cannot describe the infrared
physics in the confining phase. We therefore intend to study more
carefully down to what $N_f$ can we trust perturbation theory in the
infrared, and in particular, when does perturbation theory signal its
own inapplicability. 
The very same questions can be asked also in supersymmetric 
QCD (SQCD), where more is known about the phase structure. 
We therefore study here both QCD and SQCD and compare the two.

A few years ago Seiberg lead a revolution in the understanding of
supersymmetric gauge theories\footnote{For recent reviews see 
\cite{Peskin,Shifman_review}.}. 
Of particular interest to us is the phase
structure of $N=1$ SQCD in which non-Abelian
electric-magnetic duality plays a major role \cite{Seiberg}.   
The general arguments available in the supersymmetric case do not
apply in the absence of supersymmetry and in fact the phase structure 
of a supersymmetric theory can be quite different from that of its 
non-supersymmetric parallel \cite{Peskin,Shifman_review}. 
Still the comparison can be very enlightening.

SQCD, just like QCD, has a conformal window where the theory is
asymptotically free, and is governed by a non-trivial fixed-point in
the infrared. The picture described in \cite{Seiberg} is the following:
in the upper part of the conformal window the fixed-point value 
of the coupling 
is small, and thus the theory is weakly coupled at all scales. The massless
fields that appear in the Lagrangian conveniently describe the physics at any
scale.  
When $N_f$ becomes smaller (for a fixed $N_c$) the theory becomes strongly
coupled in the infrared. Then, it does not make sense anymore to
describe the physics in terms of the original massless
fields. Nevertheless, {\em in the infrared limit}
the theory has an effective description in terms of a dual theory:
starting with an original supersymmetric theory with an
$SU(N_c)$ gauge symmetry and $N_f$ massless chiral quark superfields
($Q_i$, $i=1,2,...N_f$) and their anti-fields ($\tilde{Q}_i$, $i=1,2,...N_f$)
 the dual theory is an $SU(N_f-N_c)$ gauge theory with $N_f$ chiral
quarks superfields
($q_i$, $i=1,2,...N_f$) and their anti-fields ($\tilde{q}_i$, $i=1,2,...N_f$), 
and an additional
superpotential describing a Yukawa interaction between color-singlet
mesons and the quarks superfields: 
\beq
W=\sqrt{\lambda}\,M_j^i\,q_i\,\tilde{q}^j.
\label{superpotential}
\eeq

The relation between the theories is referred to as duality 
since the dual of the dual theory is again an $SU(N_c)$ gauge theory.
In the conformal window the dual theory, just like the original one,
is asymptotically free and has a non-trivial infrared fixed-point.
Contrary to the original theory, the dual theory becomes weakly
coupled as $N_f$ decreases, until the point where it becomes infrared
free. Since the dual theory is weakly coupled when the original one is
strongly coupled, and vise-versa, Seiberg refers to this duality as a
non-Abelian generalization of the electric-magnetic duality.
The duality picture is valid also outside the conformal window, where
one of the theories is infrared free and the other is confining, 
but here we concentrate on the conformal window.

In \cite{Oehme_SQCD} (see also \cite{superconvergence_SQCD}) it was
shown that the lower boundary of the conformal window implied by the 
superconvergence criterion for confinement in SQCD, coincides with the
lower boundary implied by duality which is the point where the dual
theory becomes infrared free. This gives additional support to the
whole picture, with the advantage that the superconvergence criterion
can be applied also in the non-supersymmetric case, as it was
originally done in \cite{Oehme_metric_confinement,Oehme_potential_confinement}.

According to Seiberg's description, in SQCD the electric theory is strongly
coupled at the bottom of the window. We shall demonstrate that this
strong coupling behavior manifests itself already at the perturbative
level, through appearance of Landau singularities that make
perturbation theory inconsistent.

The purpose of the first part of this paper (Sec. 2) is to
consider the condition for the perturbative QCD coupling to be causal 
as a function of $N_f$ (for a general $N_c$), and compare it with
the lower boundary of the conformal window set by superconvergence.
In Sec. 2.1 we study causality at the level of the 2-loop coupling and
in Sec. 2.2 we examine the effect of higher orders.
Next, in Sec. 3, we study the same issue in SQCD. In Sec. 3.1
we investigate the singularity structure of the 2-loop coupling in the
electric theory and study the effect of higher orders. 
In Sec. 3.2 we discuss the dual (magnetic) theory. 
Sec. 3.3 summarizes the main findings of Sec. 2 and 3.
In Sec. 4 we consider the Banks-Zaks expansion for the value
of the fixed-point (Sec. 4.1) and for the critical exponent $\gamma$ 
that controls the rate at which a generic effective charge approaches 
the fixed-point in the infrared limit (Sec. 4.2). 
In Sec. 4.3 we show that the 2-point 
Pad\'e approximants technique \cite{Baker} can be used to interpolate  
between the Banks-Zaks expansions for a physical quantity in the two
dual theories. The example considered is the critical exponent
in the large $N_c$ limit.
 
\section{The conformal window in QCD and the analyticity structure of
  the coupling}

As explained in the introduction, according to the superconvergence criterion 
\cite{Oehme_metric_confinement,Oehme_potential_confinement} an
$SU(N_c)$ gauge theory with $N_f$ light flavors 
is confining so long as the anomalous dimension
of the gauge field propagator $\gamma_{00}$ of eq. (\ref{gamma00}) is
negative, i.e. for \hbox{$r=N_f/N_c<13/4$}. 
This is only a sufficient condition for
confinement, and therefore we can expect a phase transition from the
confining phase to a phase which is conformally invariant in the
infrared, either at $r=13/4=3.25$ or somewhere above this line.

Referring to the superconvergence criterion as determining the lower
boundary of the conformal window, we now turn to study the
perturbative $\beta$ function. 
In real-world QCD the perturbative running coupling has ``causality
violating'' Landau singularities on the space-like axis which,
according to the common lore, signal the inapplicability of
perturbation theory in the infrared region and the necessity of
non-perturbative power like terms. 
On the other hand, close enough to the top of the conformal window causality
can be established within the perturbative framework \cite{FP,LamW}.
There the perturbative $\beta$ function
leads to a {\em causal running coupling}, which has a finite infrared
limit and no Landau singularities in the entire $Q^2$ plane: 
its only discontinuity is a cut along the time-like axis. 
In this situation the perturbative analysis does not signal the need
for non-perturbative physics. It is then possible that 
perturbation theory by itself describes well the infrared physics. 

By definition, in the conformal window the coupling reaches a finite 
limit in the infrared. As explained above, in the upper part of the 
window this finite limit is obtainable from the perturbative $\beta$ function. 
Is it true also away from the top of the
window? In other words, is it the {\em perturbative} coupling that reaches a
finite limit? and in this case, can we reliably calculate the fixed-point
value in perturbation theory?  
In order to address these questions we study here the conditions for 
a causal perturbative coupling and compare them to the boundary of the
conformal window.

\subsection{Causality from the 2-loop $\beta$ function} 

The 2-loop $\beta$ function with $\beta_0>0$ and $\beta_1<0$ 
is the simplest example
where Landau singularities can be avoided, so it is natural to begin by
analyzing this example.
It should be stressed that the 2-loop $\beta$ function corresponds to a
particular choice of renormalization scheme, the so-called `t Hooft
scheme, where all the higher-order corrections to the $\beta$ function
$\beta_2$, $\beta_3$ and on vanish. 
Since the first two coefficients of the $\beta$
function are scheme invariant, we shall obtain a 
criterion for causality which does not have an explicit dependence 
on the scheme. 

The 2-loop renormalization group equation,
\beq
\beta(x)=\frac{dx}{dt}=-\beta_0{x}^2\left(1+cx \right)
\label{beta_2loop}
\eeq
where $t=\ln(Q^2/\Lambda^2)$, and \cite{twoloops} 
\beq
c\equiv\frac{\beta_1}{\beta_0}=\frac 1{16\beta _0}
\left[\frac{34}{3} N_c^2+\left(\frac{1}{N_c} -\frac{13}{3} N_c\right) 
N_f\right]
\label{c}
\eeq
can be integrated exactly \cite{LamW} using the Lambert W 
function \cite{Lambert}. It was shown in \cite{LamW} that 
if $c>0$ a Landau branch point is present on
the space-like axis, if $-\beta_0<c<0$ a pair of Landau branch points appears
at some complex $Q^2$ values and if
\beq
c<-\beta_0<0
\label{causality_condition}
\eeq
the coupling has a causal analyticity structure, with no Landau
singularities\footnote{We always assume asymptotic freedom, i.e. $\beta_0>0$.}.
We rederive here these results using a simpler (but less rigorous) approach 
\cite{private}.
 
Integrating (\ref{beta_2loop}) we obtain
\beq
\ln(Q^2/\Lambda^2)\,=\,
\frac{1}{\beta_0x}+\frac{1}{\gamma^{2-loop}}
\ln\left[\frac{1}{x}-\frac{1}{x_{\FP}}\right]  
\label{int_2_loops}
\eeq
where $x_{\FP}=-1/c$ and 
$\gamma^{2-loop}$ is the critical exponent at the 2-loop
order. $\gamma$ is defined as the
derivative of the $\beta$ function at the fixed
point,
\beq
\gamma\equiv\left.\frac {d\beta(x)}{dx}\right\vert_{x=x_{\FP}}.
\label{gamma_first_def}
\eeq
At 2-loop order $\gamma^{2-loop}=-\beta_0^2/\beta_1$.

If $c>0$ there is a Landau singularity on the space-like axis. A
positive fixed-point is obtained for $c<0$, but this condition alone
does not guarantee causality -- there can be Landau singularities in
the complex $Q^2$ plane.

Assuming that the singularities are such
that $\vert x(Q^2) \vert \longrightarrow \infty$, we expand
(\ref{int_2_loops}) around these points in powers of $1/x$.
The leading term in this expansion gives the location of the singularity.
The phase of the r.h.s. of (\ref{int_2_loops}) at the
singularity is  
\beq
\Phi\,=\,\pm \pi\,\frac{1}{\gamma^{2-loop}},
\eeq
If $\vert\Phi\vert<\pi$, i.e. if $\gamma^{2-loop}>1$, or $-\beta_0<c<0$, 
the singularities are in the first sheet 
(the time-like axis cut corresponds to $\vert\Phi\vert=\pi$), and if 
$\vert \Phi \vert>\pi$, i.e. if
\beq
0<\gamma^{2-loop}<1
\label{2_loop_causality}
\eeq
or $c<-\beta_0$, the 2-loop coupling is causal. 
The condition (\ref{2_loop_causality}) (or (\ref{causality_condition}))
for a causal 2-loop coupling translates in QCD
into the following condition for $r$ upon substituting $\beta_0$ and
$c$ from eqs. (\ref{beta0}) and (\ref{c}), respectively:
\beq
4r^2+\left[\frac{9}{N_c^2}-83\right]r+223<0
\label{r_QCD_eq}
\eeq
This leads to an approximately $N_c$ independent critical value for
$r$ for any possible value of $N_c$ (since $N_c^2\gg 9/83$), namely the
2-loop coupling is causal as long as 
\beq
r>(83-9\sqrt{41})/8\simeq 3.17.
\label{r_QCD}
\eeq

For lower $r$, the condition (\ref{2_loop_causality}) does not hold
and there appears a pair of complex singularities in the $Q^2$
plane. If $r$ is reduced further, $c$ becomes positive and then a 
Landau branch point appears on the space-like axis. This change occurs at:
\beq
r=34/\left[13-\frac{3}{N_c^2}\right]\simeq 2.62.
\label{r_space_like_QCD}
\eeq

The results are summarized in fig.~\ref{conformal_window} in the upper plot,
where the lower boundary of the
conformal window implied by superconvergence ($r=3.25$) is compared
with the lower boundary of the region where the 2-loop coupling is causal
according to (\ref{r_QCD_eq}), which is asymptotic at large $N_c$ to
$r\simeq 3.17$. 
Clearly, the {\em 2-loop coupling is causal in the entire conformal window}. 
This conclusion holds, of course, also in the case
where the lower boundary of the conformal window is somewhere above the
critical value for superconvergence ($r=3.25$).   
This suggests that the fixed-point in QCD is always of perturbative origin.

The proximity of the two lines, the upper boundary of the 
superconvergence region ($r=3.25$) and the lower boundary of the 
2-loop causality region ($r=3.17$) does not have any deep
meaning. Presence of complex Landau singularities in the running
coupling signals that the coupling becomes strong 
but it does not necessarily imply confinement -- an example is provided by SQCD
(Sec. 3). 

Due to the closeness of the two lines one might worry that 
even within the conformal window the large distance physics cannot 
be reliably described by perturbation theory. However, we shall see in the next
section that 3-loop corrections make the coupling causal in a wider
range, and eventually perturbation theory does seem reliable down to
the bottom of the conformal window.  

\subsection{How relevant is the criterion for causality at 2-loop?}

It is natural to wonder whether the singularity structure of the
coupling which is defined by the
truncated 2-loop $\beta$ function is of any physical significance. 
Of course we do not doubt the assumption that the theory as a whole is
causal. According to the common lore, the appearance of the 
non-physical Landau singularity in the perturbative coupling in
real-world QCD is nothing more than a sign of the inapplicability of 
perturbation theory for describing the infrared region. 
Thus the presence of Landau singularity indicates 
the significance of non-perturbative corrections in the infrared.
The interesting point is that close enough to the top of 
the conformal window, there may be a
possibility to {\em establish} causality using only perturbation
theory, as we explain below. 

\subsubsection{Causality beyond 2-loop -- general discussion}

In general, the analyticity structure of a coupling based on some
higher order $\beta$ function, 
\beq
\beta(x)=-\beta_0x^2\left[1+cx+c_2x^2+\cdots\right]
\label{beta_function}
\eeq
can be completely different from that of
the 2-loop coupling (\ref{beta_2loop}). 
This is clearly so if Landau singularities
are present: their location and nature generally depend on
{\em all} the coefficients of the $\beta$ function and consequently on the
renormalization scheme. This ``instability'' should be of no surprise
since the weak coupling expansion breaks down completely when
examining the singularities of the coupling.

As an example how the singularity structure changes and becomes
more complex as higher order terms in the $\beta$ function are 
included, consider the 1-loop coupling, the 2-loop coupling and 
Pad\'e improved 3-loop coupling, defined by 
\beq
\beta_{\PA}(x)=-\beta_0x^2\frac{1+[c-(c_2/c)]x}{1-(c_2/c)x},
\label{PA}
\eeq
which were all analyzed in \cite{LamW}. The 1-loop coupling has a
space-like Landau pole, the 2-loop coupling can have a causal
structure or a pair of complex branch points or a space-like branch
point. The Pad\'e improved 3-loop coupling can be causal but it can
also have both simple poles and branch points (the details appear 
in \cite{LamW}). 
 
While these examples show that there is no stability when going to
higher orders if Landau singularities exist, they also indicate that if
the 2-loop coupling is causal, causality may be preserved when
higher order corrections are included. In fact, it is rather simple to
explain why this kind of stability be expected in general.
When the 2-loop coupling is causal it is bounded, and in many
cases also small, for any complex $Q^2$. If so the usual
perturbative justification holds: the next term in the $\beta$
function series which is
proportional to a higher power of the coupling is small, 
and likewise higher order terms. In this situation higher-order terms
are not expected to have much influence on $x(Q^2)$.  
In other words, {\em absence} of Landau singularities can be
consistently confirmed at the perturbative level, whereas {\em
presence} of Landau singularities can only be confirmed or disproved
in the full theory by non-perturbative  methods. 

The first step in establishing causality in perturbation theory
is to examine the analyticity structure of the 2-loop coupling, as we
did above. On one hand the 2-loop coupling 
has the advantage that it does not depend 
on the renormalization scheme. On the other hand it does not
correspond directly to any observable and therefore it may not be causal. 
Thus, we are forced to examine higher order corrections (or
renormalization schemes other than the `t Hooft scheme), and see 
whether the causality condition at 2-loop order is reasonable.
The next step is therefore to choose a representative renormalization
scheme, different from the `t Hooft scheme, and ask whether the
3-loop correction to the $\beta$ function has a significant effect 
on the infrared coupling. If the effect of the 3-loop correction is
negligible, that is if the coupling is small enough such that  
\beq
\left\vert \beta_2x^2(Q^2)\right\vert\ll \left\vert \beta_1 x(Q^2)\right\vert
\label{causality_stable_2loops}
\eeq
in the {\em entire complex $Q^2$ plane}, then we shall consider 
that causality is established at the perturbative level. 

To be completely convinced, one might want to check also the magnitude of 
higher-order corrections corresponding to 4-loop order and beyond.
However, it is important to remember in this respect, that if we go to
high enough order ($n$), we will always obtain 
\beq
\left\vert \beta_nx^n\right\vert> \left\vert \beta_{n-1}x^{n-1} \right\vert
\eeq
due to the asymptotic nature of the $\beta$ function series, and thus
it does not make sense to require that {\em all} the higher-order 
terms will be small. 
In the scenario described above, namely that the 2-loop coupling is
already causal and small it seems reasonable to require that 
the 3-loop correction is small and stop there.
Clearly, this scenario is just the simplest case to consider. It is
possible that $x(Q^2)$ at 2-loop is still not small enough so as to guarantee
$\vert \beta_2 x^2\vert\ll \vert \beta_1 x\vert$, but  
$\beta_2$ is negative so  
$x(Q^2)$ at 3-loop is much smaller, and then higher order corrections
are negligible: $\vert \beta_3x^3\vert \ll \vert \beta_2x^2\vert$. 
Of course, in this case
the results might depend on the renormalization scheme. 

An encouraging observation with regards to the 2-loop analysis 
is that the condition for causality of
the 3-loop coupling is quite modest once the 2-loop coupling is
causal. We will show that the only further requirement is that the 3-loop
$\beta$ function has a positive root corresponding to the infrared 
stable fixed-point.
 
It is most convenient for this demonstration to write the 3-loop
$\beta$ function in the following form:
\beq
\beta(x)\equiv\frac{dx}{d\ln(Q^2)}=-\beta_2 x^2
\left[f_1f_2-(f_1+f_2)x+x^2\right]
\label{beta_roots}
\eeq
where  $f_1f_2=\beta_0/\beta_2$ and $f_1+f_2=-\beta_1/\beta_2$, thus:
\begin{eqnarray}
f_1&=&\frac{1}{2\beta_2}\left(-\beta_1+\sqrt{\Delta}\right)
\\ \nonumber
f_2&=&\frac{1}{2\beta_2}\left(-\beta_1-\sqrt{\Delta}\right)
\\ \nonumber
\label{f_12}
\end{eqnarray}
with $\Delta=\beta_1^2-4\beta_0\beta_2$.
Note that in all cases of interest, namely when there is a positive real
zero to the 3-loop $\beta$ function, the infrared fixed-point is
$x_{\IR}\equiv f_2>0$, and $x_{\UV}\equiv f_1$ is an ultraviolet fixed-point.
The corresponding critical exponents are, for $x_{\IR}$:
\beq
\gamma_{\IR}^{3-loop}
=\beta_2x_{\IR}^2(x_{\UV}-x_{\IR})=x_{\IR}^2\sqrt{\Delta}>0
\label{gamma_2_3_loops}
\eeq
and for $x_{\UV}$:
\beq
\gamma_{\UV}^{3-loop}
=-\beta_2x_{\UV}^2(x_{\UV}-x_{\IR})=-x_{\UV}^2\sqrt{\Delta}<0.
\label{gamma_1_3_loops}
\eeq
It is useful to note that
\beq
\frac{1}{\gamma_{\UV}^{3-loop}}+\frac{1}{\gamma_{\IR}^{3-loop}}
\,=\, \frac{1}{\gamma^{2-loop}}\, =\, -\frac{\beta_1}{\beta_0^2},
\eeq
where we have used the definition of $f_{1,2}$. It then follows,
assuming 2-loop causality (\ref{2_loop_causality}), that
\beq
0<\gamma_{\IR}^{3-loop}<\gamma^{2-loop}<1.
\label{gamma_ineq}
\eeq

In order to examine causality at 3-loop we integrate
(\ref{beta_roots}) and obtain:
\beq
\ln(Q^2/\Lambda^2)\,=\,
\frac{1}{\beta_0x}+\frac{1}{\gamma_{\UV}^{3-loop}}
\ln\left[\frac{1}{x}-\frac{1}{x_{\UV}}\right] 
+\frac{1}{\gamma_{\IR}^{3-loop}}\ln\left[\frac{1}{x}-\frac{1}{x_{\IR}}\right] 
\label{int_3_loops}
\eeq
To find the causality condition, we study, as in the 
2-loop case, the phase of the Landau singularity\footnote{As 
opposed to the 2-loop case, where it is also possible to invert
\cite{LamW} the relation (\ref{int_2_loops}) to
calculate $x(Q^2)$ using the Lambert W function, 
here this cannot be done.}. 
We assume that the only singularities are such
that $\vert x(Q^2) \vert \longrightarrow \infty$, and expand
(\ref{int_3_loops}) around these points in powers of $1/x$. The leading
term in this expansion gives the location of the singularity.
If $\beta_2<0$, then $x_{\UV}$ is negative and the phase of the 
r.h.s. of (\ref{int_3_loops}) at the singularity is  
\beq
\label{Phi_3_loop_1}
\Phi\,=\,\pm \pi\,\frac{1}{\gamma_{\IR}^{3-loop}},
\eeq
while if $\beta_2>0$, $x_{\UV}$ is positive and the phase is  
\beq
\label{Phi_3_loop_2}
\Phi\,=\,
\pm \pi\,\left(\frac{1}{\gamma_{\UV}^{3-loop}}
+\frac{1}{\gamma_{\IR}^{3-loop}}\right)
\,=\pm\,\pi\, \frac{1}{\gamma^{2-loop}}.
\eeq  
Using (\ref{gamma_ineq}) we find that in
both cases $\vert\Phi\vert>\pi$  and it follows that the 
3-loop coupling is causal. 

We showed that if the 2-loop coupling is causal, the 3-loop coupling is
also causal, provided it has an infrared fixed-point. It is not clear
whether such a conclusion can be extended to higher orders. 
It may be however interesting to note that we already know
from the analysis of \cite{LamW} another example where similar
conclusions hold: this is the Pad\'e improved 3-loop $\beta$ function,
defined by (\ref{PA}). Contrary to the above examples, this $\beta$
function is not truncated at some finite order, and thus it could be
expected a priori to behave differently.  
According to \cite{LamW} the causality condition for the 
Pad\'e-improved 3-loop coupling
is $c<-\beta_0$ and $c_2<c^2$. The first condition is
the same as the condition for the causality of the 2-loop coupling.
In fact, the critical exponent of this coupling is equal to that of
the 2-loop coupling, and thus the condition is 
$0<\gamma^{\PA}=\gamma^{2-loop} <1$.
The second condition is just the condition to have a positive infrared
fixed-point.

We comment that the inverse statement does not hold: the 3-loop
coupling can be causal even if the 2-loop coupling is not.
If $\beta_2$ is negative and large enough, the 3-loop coupling 
is causal independently of the sign of $\beta_1$.
  
Coming to analyze the conditions for causality or the stability of
the causal solution with respect to higher order corrections (such as 
(\ref{causality_stable_2loops})), we should, in general solve the
renormalization group equation at each order to obtain  
$x(Q^2)$ in the entire $Q^2$ plane, as was done in \cite{LamW} for the
2-loop and the Pad\'e improved 3-loop couplings.
However, we shall demonstrate below that it is in fact enough to examine the
effect of higher orders on the
infrared limit of the {\em space-like}
coupling $x_{\FP}\equiv x(0)$, unless the coefficients of
the $\beta$ function are {\em extremely} close to the condition where
causality is lost.
In most cases when $x(Q^2)$ is causal, 
$\vert x(Q^2)\vert\lsim x(0)$ in the entire $Q^2$ plane. 
Of course, when causality is lost 
$x(Q^2)$ diverges at some point while $x(0)$ is finite, 
and thus close to the boundary of the
causality region $x(0)$ is not indicative at all. The point is that 
the region where the maximum value of $\vert x(Q^2)\vert$ is much
larger than $x(0)$ is quite narrow. 
To demonstrate this, consider again the example of 2-loop coupling 
in $N_c=3$ QCD. According to (\ref{r_QCD_eq}) this coupling is causal so
long as $N_f\gsim 9.683$. At the point where causality is lost,
$\vert x(Q^2) \vert$ reaches infinity on the first sheet (on the
time-like axis), while the space-like coupling has its maximum at 
$x(0)\simeq 0.88$ which is not so large. However, if we move slightly
above the causality boundary, the maximal value of $\vert x(Q^2) \vert$ 
in the entire $Q^2$ plane becomes of the order of $x(0)$. We show this
phenomenon in fig.~\ref{x_plane_Nc3} where we plot the region in the
complex coupling plane, into which the entire complex $Q^2$ plane is
mapped. The contour itself corresponds to the cut along the time-like
axis ($Q^2<0$) and it was computed using the Lambert W function
solution, as explained in \cite{LamW,FP}. As shown in the plot, for
$N_f=9.7$, i.e. very close to the point where causality is lost, the
maximal value of $\vert x(Q^2)\vert$ on the time-like axis 
is still significantly
larger than $x(0)$. One clearly identifies here the effects of the
singularities that are present on the second sheet. On the other hand,
already at $N_f=10$, the maximal value of $\vert x(Q^2)\vert$ 
on the time-like axis is of the order of $x(0)$.

We found that the condition for the 2-loop coupling to have a causal
analyticity structure is $0\leq\gamma^{2-loop}<1$, where
$\gamma^{2-loop}=0$ corresponds
to a free theory, the limit obtained at the top of the conformal
window, and $\gamma^{2-loop}=1$ corresponds to the point where Landau
singularities first appear. At 2-loop order the condition
$0\leq\gamma^{2-loop}<1$ is both a sufficient and a necessary condition. It is
interesting to see how this generalizes to higher orders. 
When the $\beta$ function has more than one zero, we should specify at
which of them $\gamma$ is defined. The only root that is relevant
in the asymptotically free phase is the smallest
positive zero, the physical infrared stable fixed-point, 
and we always refer to this one.

The 3-loop analysis shows that $0\leq\gamma^{3-loop}<1$ is a necessary
condition but not a sufficient one. An example where the 3-loop
coupling is not causal although the above condition is obeyed can be
constructed starting with a non-causal 2-loop $\beta$ function with
$\beta_1<0$ and adding a 3-loop term with $\beta_2$ positive but small
enough such that a positive zero for the
3-loop $\beta$ function exists. It then follows from eq. (\ref{Phi_3_loop_2})
that the 3-loop coupling is not causal although $\gamma^{3-loop}$ can
still obey the above condition. We stress that this example is
not representative since usually, as we shall see, $\beta_2<0$ and then
the condition $0\leq\gamma<1$ is both necessary and sufficient 
also at the 3-loop order.

In fact the condition $0\leq\gamma<1$ is {\em always} necessary for a causal
analyticity structure.
The condition $\gamma\geq 0$ is simply the one to have an
infrared stable fixed-point.  
To show that also $\gamma<1$ is necessary we use the following observation:  
a causal structure implies that there is a well defined mapping $x(Q^2)$
from the entire complex $Q^2$ (the first sheet) into a 
{\em compact domain} in the complex coupling plane, such that for large
enough $\vert Q^2 \vert$ the coupling flows to the trivial
fixed-point, as implied by asymptotic 
freedom\footnote{It was demonstrated in \cite{LamW} in the
particular case of the Lambert W solution for the 2-loop coupling that in order
to define the analytical continuation of $x(Q^2)$ from the space-like
axis to the entire first sheet, it is essential to require asymptotic
freedom for complex $Q^2$ values.}.
As we saw in the example of fig.~\ref{x_plane_Nc3} (these features as
completely general) the space-like axis
is mapped to real positive values in the range $[0,x_{\FP}]$ and the
time-like axis is mapped to the boundary of this domain in the complex
coupling plane. 
It follows from the definition of $\gamma$ in
(\ref{gamma_first_def}) that the coupling approaches the fixed-point
according to
\beq
\label{first_x_x_fp}
x=x_{\FP}-\left(\frac{Q^2}{\Lambda_{\eff}^2}\right)^{\gamma},
\eeq
where $\Lambda_{\eff}$ is an observable-dependent QCD scale.
If $\gamma>1$, there is a phase $\Phi=\pi/\gamma$ in the
complex $Q^2$ plane ($Q^2=Q_0^2\exp(i\Phi)$) such that in the limit
$Q^2_0\longrightarrow 0$ the rays $\pm\Phi$ are mapped by
(\ref{first_x_x_fp}) to positive real
values of the coupling {\em larger} than the fixed-point value
($x=x_{\FP}^+$). On the other hand a straightforward analysis of the
$\beta$ function shows that values of the coupling $x>x_{\FP}$ either 
belong to the domain of attraction of some {\em non-trivial} ultraviolet
fixed-point or flow to an ultraviolet Landau singularity. 
The conclusion is that there is no singularity free mapping that obeys
the asymptotic freedom condition stated above. In particular, if two different
ultraviolet fixed-points are allowed for different values of $Q^2$ it
implies the existence of a separatrix, discriminating between the values
of $Q^2$ that flow to each of the ultraviolet fixed-points, i.e. there
are singularities in the first sheet of the complex $Q^2$ plane.
We stress that the arguments why $0\leq \gamma<1$ is a necessary condition
for a causal analyticity structure are completely general: they are
not based on perturbation theory.

\subsubsection{Causality at higher orders in QCD}

We would like to examine whether causality can be established
in perturbation theory in the specific case of the conformal window in
QCD. Close to the top of the conformal window, causality is established
at the 2-loop level. The infrared coupling is small and thus the
3-loop term is negligible and 
condition (\ref{causality_stable_2loops}) 
for stability of the perturbative analysis is obeyed. This is no longer true at
the bottom of the window.

We start the discussion in the ${\hbox{$\overline{\hbox{MS}}\,$}}$
scheme, which has the advantage that the
4-loop coefficient in the $\beta$ function is known 
\cite{threeloops,fourloops}.
We shall refer to physical effective charges later.
The fixed-point value of the coupling, calculated as an
explicit solution of the equation $\beta(x)=0$ in the large $N_c$
limit at 2-loop and then in the ${\hbox{$\overline{\hbox{MS}}\,$}}$ 
scheme at 3-loop and 4-loop 
orders is shown in fig.~\ref{QCD_IR_epsilon} as a function of the 
distance from the top of the conformal window,  
\beq
\epsilon\equiv\frac{11}{2}-r=\frac{11}{2}-\frac{N_f}{N_c}. 
\eeq
The 2-loop coupling reaches relatively large
values towards the bottom of the conformal window, but then the 3-loop
and 4-loop couplings take significantly lower values, and in addition
they are very close to each other. These results can be understood
knowing the negative sign of the 3-loop coefficient $\beta_2$ 
and the magnitude of successive terms in the
${\hbox{$\overline{\hbox{MS}}\,$}}$ scheme, shown in fig.~\ref{QCD_terms_}. 
In the latter, the coupling is evaluated at the fixed-point 
according to the zero of the 3-loop $\beta$ function.
It is clear from the plot that the condition for stability of the 2-loop result
(\ref{causality_stable_2loops}) does not hold in the lower part of the
conformal window. It certainly does not hold for $r \lsim 4$,
corresponding to $N_f\lsim 12$ since there the 3-loop term is comparable to
the 2-loop term. 
Thus we are forced to examine causality at higher orders. 

Since the 3-loop coefficient in ${\hbox{$\overline{\hbox{MS}}\,$}}$
is negative for the relevant $N_f/N_c$ values,
the 3-loop $\beta$ function has a positive real fixed-point, and
according to the general discussion in the previous section, 
3-loop causality is
implied within the region where the 2-loop coupling is causal.
Now, in order to trust the 3-loop causality, it is required that the
4-loop term will be small enough. Indeed, as shown in fig.~\ref{QCD_terms_}
the 4-loop term in the ${\hbox{$\overline{\hbox{MS}}\,$}}$ scheme 
remains small in the
entire conformal window. The effect of the 4-loop term on the
fixed-point value is shown in fig.~\ref{QCD_IR_epsilon}. Clearly,
this is a negligible effect, and thus perturbative stability is
realized at the 3-loop level.
It would be better to check the effect of the
4-loop term on $x(Q^2)$ in the entire $Q^2$ plane, but based on the experience
with the 2-loop coupling we expect that in general 
the space-like fixed-point value is
indicative of the magnitude of $x(Q^2)$ in the entire complex $Q^2$
plane. 

Next we consider the value of the critical exponent as a function of
the distance from the top of the window. The results of an explicit
calculation of $\gamma$, in the large $N_c$ limit, 
from the 2-loop, 3-loop and 4-loop $\beta$
functions in ${\hbox{$\overline{\hbox{MS}}\,$}}$ are shown in 
fig.~\ref{combined_gamma_BZ} in the upper plot. In agreement with our
previous discussion the condition 
\hbox{$0<\gamma^{3-loop}<\gamma^{2-loop}<1$} is obeyed in the entire
conformal window. The points where the 2-loop and 3-loop couplings
cease to be causal can be identified in this figure as the points where
$\gamma=1$. Since the 4-loop term is small,
$\gamma^{4-loop}\simeq\gamma^{3-loop}$ within the resolution of this
plot, and so the perturbative
stability which characterises the coupling exists also for the critical
exponent.  
  
We stress that the results described above 
are not special to the large $N_c$ limit. In particular
fig.~\ref{QCD_IR_epsilon} through \ref{combined_gamma_BZ} are 
qualitatively the same for any $N_c$.

The above investigation shows that the ${\hbox{$\overline{\hbox{MS}}\,$}}$
coupling is causal at the 3-loop level and, given the smallness of the
4-loop term, presumably also at the 4-loop level in the entire 
conformal window. However, this coupling does not correspond
directly to any observable quantity. It is important to check whether
similar conclusions apply in physical schemes.   

A relevant analysis has been performed in \cite{FP}. Fig.~1 in
\cite{FP} compares the $N_f$ dependence of $c_2\equiv \beta_2/\beta_0$
for various physical effective charges. The observation that 
$c_2$ for different effective charges are numerically close and that
they share the same $N_f$ dependence indicates that certain
properties of the coupling may be generic in spite of scheme dependence.  
In particular we note that $c_2$ is negative in the entire conformal window
not only in ${\hbox{$\overline{\hbox{MS}}\,$}}$, but also for all the
physical effective charges considered. 
We conclude that there is a fixed-point at the 3-loop
order in all these physical schemes and, according to the general discussion
above, 3-loop causality follows\footnote{Note that the Pad\'e improved 3-loop
coupling is also causal.}. Unfortunately, 4-loop coefficients in
physical renormalization schemes are not known yet\footnote{An
exception is the effective charge related to the Higgs hadronic decay width.
For this quantity the infrared fixed-point does not even exists in the
lower part of the conformal window due to a large positive 4-loop coefficient. 
We do not, however, consider this example as representative (see the
discussion in \cite{FP}).}. 
Consequently, the stability of the 3-loop causal coupling with respect
to higher loop corrections cannot be studied for physical effective
charges like we did in the ${\hbox{$\overline{\hbox{MS}}\,$}}$ scheme. 
However an alternative is provided by the Banks-Zaks expansion, 
which can be calculated in physical schemes up to
next-to-next-to-leading order term \cite{BZ_grunberg,CaSt,FP}. 
This will be discussed further in Sec. 4.     

We comment that the perturbative coupling at the 3-loop order can have
a causal analyticity structure even somewhat below the bottom of the conformal
window, i.e. in the upper part of the confining phase. 
In this respect, different couplings may behave differently.
We recall that for $N_c=3$ the `t Hooft coupling, defined by the truncated
2-loop $\beta$ function, is causal down to $N_f\simeq 9.68$, quite
close to the bottom of the conformal window $N_f\simeq 9.75$.  
This can be compared with the ${\hbox{$\overline{\hbox{MS}}\,$}}$ 
scheme where 3-loop causality is lost at $N_f\simeq 8.5$ and to
physical renormalization schemes in which the causality domain is even
wider. Based on the results of \cite{FP} and the above type of
analysis we find that for the effective charge defined by the 
vacuum polarization D-function and the ones associated with
the polarized and non-polarized Bjorken sum-rules 3-loop causality is lost at
$N_f\simeq 7.2$ while for the effective charge defined from the heavy
quark potential 3-loop causality is lost at 
$N_f\simeq 8.4$\footnote{The last result is based on the recently published 
2-loop calculation of the static potential in QCD
\cite{V-scheme-corrected}, which corrects a previous result used in 
\cite{FP}.}.

Finally we consider the calculation of the critical exponent using
physical renormalization schemes. As long as the fixed point is
perturbative, it is natural to expect that $\gamma$ could be 
calculated with a reasonable accuracy starting with the truncated 
$\beta$ function in various renormalization schemes.   
Since $\gamma$ is a universal quantity the results should agree.
The results of an explicit calculation of $\gamma$ in several physical schemes 
at the 3-loop order are presented in fig.~\ref{combined_gamma_BZ}
together with the results in ${\hbox{$\overline{\hbox{MS}}\,$}}$. 
The schemes we use include the vacuum polarization D-function, the
polarized and non-polarized Bjorken sum-rules (the latter two curves
overlap) and the heavy quark effective potential.
The results in the different schemes agree very well close to 
the top of the window. The spread increases to about $\pm 15 \%$ 
towards the bottom of the window and is interpreted as an artifact of 
using a truncated perturbative expansion. We shall come back to
discuss the accuracy to which $\gamma$ can be calculated in sec. 4.2.1
in the framework of the Banks-Zaks expansion (see table 5 there).

\section{The conformal window in SQCD}

\setcounter{footnote}{0}
The $\beta$ function in SQCD is given by\footnote{Capital letters 
are used here to distinguish SQCD coefficients from QCD ones.}
\beq
\beta(x)\,\equiv\,\frac{dx}{d\ln(Q^2)}
\,=\,-(B_0x^2+B_1x^3+\cdots)\,=\,-B_0x^2\,(1+C_1 x+\cdots)
\label{beta_SQCD}
\eeq 
where $x=\alpha/\pi=g^2/(4\pi^2)$,
\beq
B_0= \frac14 \left(3N_c-N_f\right),
\label{B0}
\eeq
and 
\beq
C_1\equiv \frac{B_1}{B_0}= \frac12N_c-\frac{N_f}{3N_c-N_f}\frac{N_c^2-1}{2N_c},
\label{C1}
\eeq
where the coefficients where calculated in \cite{NSVZ}.
Above the line $R\equiv N_f/N_c=3$ the theory is infrared free, while below
this line it is asymptotically free in the ultraviolet.
For $R$ just below $3$, $B_0$ is small and positive and $B_1$ is
negative, leading to an infrared fixed-point at a small 
$x_{\FP}\simeq -B_0/B_1=-1/C_1$, making the theory weakly coupled at all
scales. As $N_f$ (and thus $R$) is decreased, the infrared coupling
increases. According to Seiberg \cite{Seiberg}, the infrared
fixed-point persists even down to such low $N_f$ that the 
original degrees of freedom are strongly coupled 
and then a dual theory which is based 
on another gauge group with $N_c^d=N_f-N_c$ colors is appropriate to
describe the infrared limit
($d$ stands for a dual variable). Seiberg's conjecture can only
be understood if the fixed-point is of {\em non-perturbative } origin,
at least in the lower part of the conformal window.
This is contrary to our previous observation concerning the
{\em perturbative} origin of the fixed-point in the non-supersymmetric
case. Thus we would like to {\em check} that indeed a definite difference
exists between the conformal window in QCD, which is perturbative and
the one in SQCD which is not. This is done here by considering
the analyticity structure of the coupling constant and in the next
section, by comparing the Banks-Zaks expansion in SQCD to that in QCD.
We shall indeed see that already at the perturbative level SQCD is 
more strongly coupled than QCD in the lower part of the conformal
window. 
The fact that the strong coupling nature of SQCD at the lower part 
of the window is manifested in perturbation
theory is not obvious a priori; strong infrared effects 
could have been induced instead by terms invisible to perturbation theory.

Duality \cite{Seiberg} provides an intuitive description of the
conformal window in SQCD, which is absent in QCD. The lower boundary
of the conformal window in SQCD is naturally identified as the
$R\equiv N_f/N_c$ ratio at which the {\em dual} theory undergoes a 
phase transition
from the asymptotically free phase (inside the window) to the infrared
free phase (below the window). The 1-loop $\beta$ function coefficient 
in the dual theory can be obtained by substituting $N_f-N_c$ for $N_c$
in (\ref{B0}): 
\beq
B_0^d= \frac14\left(2N_f-3N_c\right).
\label{B0d}
\eeq
Thus, the conformal window is $3/2<R<3$, as shown in the lower plot of
fig.~\ref{conformal_window}. 
The original theory is weakly coupled, and therefore provides a 
natural physical description, for $R$ just below the line $R=3$, while the dual
theory is weakly coupled just above the line $R=3/2$. 

An important consistency check for both duality and the
superconvergence criterion for confinement is that the lower boundary
of the conformal window in both approaches coincides \cite{Oehme_SQCD}\footnote
{The generalization of this result to other supersymmetric models was
examined 
in \cite{superconvergence_SQCD}.}. The observation of \cite{Oehme_SQCD} is the 
following: 
in SQCD the anomalous dimension of the gluon propagator in the Landau gauge is 
\beq
\gamma_{00}^{\rm SQCD}=-\frac14 \left(\frac32N_c-N_f\right),
\label{Gamma00}
\eeq
which is just proportional to the first coefficient of the $\beta$
function $B_0^d$ {\em in the dual theory}. As a result,
$\gamma_{00}^{\rm SQCD}$
becomes negative, implying superconvergence and therefore confinement 
for the original theory, as $R$ becomes smaller than $3/2$, i.e. 
{\em exactly} where the dual theory becomes infrared free ($B_0^d$ in
(\ref{B0d}) changes sign).     

\subsection{The analyticity structure of the SQCD coupling}

The purpose of this section is to analyze the singularity structure of
the perturbative SQCD coupling, in parallel with the analysis of the 
QCD coupling in Sec. 2, and in particular to find when it is consistent with
causality. 

The first step is to analyse the 2-loop coupling. 
The 2-loop causality condition $C_1<-B_0$ translates, using eqs. (\ref{B0}),
and (\ref{C1}), to the following condition for $R$:
\beq
R^2+\left[\frac{2}{N_c^2}-10\right]R +15<0.
\label{R_SQCD_eq}
\eeq
Similarly to the non-supersymmetric case (see eq. (\ref{r_QCD_eq})),
the condition (\ref{R_SQCD_eq}) leads to an approximately $N_c$ independent 
critical value for $R$ for any possible value of 
$N_c$ (since $N_c^2\gg 2/10$), namely the 2-loop coupling is causal as long as
\beq
R>5-\sqrt{10}\simeq 1.8377.
\label{R_SQCD}
\eeq

The crucial observation is that the line (\ref{R_SQCD}) that limits
from below the region where the 2-loop coupling is causal, is {\em within} the
conformal window which has its lower boundary at $R=3/2$. This is
shown in the lower plot of fig.~\ref{conformal_window}.
The situation encountered here is contrary to the one in non-supersymmetric 
QCD, where the 2-loop perturbative coupling is causal in the entire 
conformal window. 
This observation fits the general expectation based on duality, that the
fixed-point in SQCD is non-perturbative in the lower part of the
conformal window. 

In addition we ask when does the 2-loop coupling develop a space-like
Landau singularity. The condition $C_1>0$ translates (using
(\ref{C1})) to the following:
\beq
R<\frac{3}{2-(1/N_c^2)},
\label{r_cd}
\eeq
which is asymptotic in the large $N_c$ limit to the lower boundary of
the conformal window, $R=3/2$.

Note that in the supersymmetric case, it is natural to use the
NSVZ form \cite{NSVZ,SV} of the $\beta$ function, and thus one may 
wonder if our
results concerning the analyticity structure of the coupling may vary
when using the $\beta$ function in this form rather than the truncated
2-loop one. In the Appendix we show that both the condition for
a causal coupling and the condition for a space-like Landau
singularity are exactly the same in the two cases, if in the NSVZ form
one uses the leading order approximation for the matter
field anomalous dimension.

The next step in the analysis of the perturbative coupling causality, 
as in the QCD case, should be to examine the effect of
higher order terms in the perturbative $\beta$ function. We choose to
work in the DRED renormalization scheme \cite{DRED}, assuming that our
conclusions will not depend on this choice.

The explicit solutions of $\beta(x)=0$ for $N_c x(0)$ in the large
$N_c$ limit are shown in fig.~\ref{SQCD_IR_delta} as a function of the
distance from the top of the conformal window,
\beq
\delta\equiv 3-R=3-\frac{N_f}{N_c}.
\eeq
The 2-loop solution is infinite at the bottom of the window (see
(\ref{r_cd})). Already here we encounter a situation different from
QCD, namely {\em stronger coupling}.   
Since the 3-loop coefficient is negative, the 3-loop solution is
smaller.  The latter is finite down to the
bottom of the window, but it is still rather large.
The fixed-point at 4-loop order exists only up to
$\delta\simeq 0.4$ (near the 4-loop arrow in the figure).
Beyond this point there is no positive real solution to the equation 
$\beta(x)=0$. The reason is that the 4-loop term in SQCD is
positive (like in QCD) and large (contrary to QCD) as can be learned
from fig. \ref{SQCD_terms}.
This figure shows the relative magnitude of the four leading
terms in the large $N_c$ SQCD $\beta$ function. 
The coupling in fig.~\ref{SQCD_terms} is evaluated 
as the zero of the 3-loop $\beta$ function.

In the lower plot of fig.~\ref{combined_gamma_BZ} we show the value of
the critical exponent as a function of $\delta$ according to the
2-loop, 3-loop and 4-loop order large $N_c$ DRED $\beta$ function. 
The necessary condition for a causal structure $\gamma=1$ 
is reached by both the 2-loop order, which was discussed
above, and 3-loop order solutions for $\gamma$ well within the
conformal window. The 4-loop result for $\gamma$ exists of course only
up to $\delta\simeq 0.4$ where a positive fixed-point exists. 
 
Examining fig.~\ref{combined_gamma_BZ}  through \ref{SQCD_terms} 
we can determine where causality can be established in SQCD 
at the perturbative level. 
In the upper part of the conformal window the 3-loop term is
small with respect to the 2-loop one, so one can trust 2-loop
causality.  As $R$ is decreased the 3-loop term becomes comparable to
the 2-loop term and then one
has to consider causality at 3-loop order.  
The negative sign of $\beta_2$ guarantees that the 3-loop
coupling is causal at least as long as the 2-loop coupling is.
But since the 4-loop term is very large, the perturbative 
argumentation fails. 
Thus in SQCD, it is possible to establish causality in 
perturbation theory {\em only in the upper part of the conformal
window}.
To be specific, two alternative criterions can be considered:
the first is to require that the 3-loop term will be smaller than 
the 2-loop term. The two become equal around $R\simeq1.9$, 
i.e. just above the 2-loop causality boundary. The second 
is even more restrictive, namely to require that also the 4-loop
term is small, or that the 4-loop $\beta$ function will have a positive
real root. This is realized only above $R\simeq 2.6$.

Maybe the most interesting observation in fig.~\ref{SQCD_terms} is the
fact that the 4-loop term in the SQCD $\beta$ function is larger than
the leading terms already very close to the top of the conformal
window. This may be related to the asymptotic nature of the
$\beta$ function series.   
The asymptotic behavior is another aspect in which the SQCD $\beta$ 
function is presumably different
from the QCD one, a point which certainly deserves further study.

\subsection{Reduction of Couplings in the magnetic theory} 

In the previous section we studied the singularity structure of the
2-loop coupling in the electric theory. Our aim here is to perform a
parallel analysis in its dual, the magnetic theory. This is, however,
not straightforward since the magnetic theory has {\em two} couplings,
rather than one. The running of the gauge coupling is affected by the
Yukawa interaction of the chiral quark superfields with the mesons,
which is described by the superpotential (\ref{superpotential}).
This gives rise to coupled renormalization group equations of the form
\begin{eqnarray}
\label{coupled_RG}
\beta^d_x(x,\lambda)\equiv \frac{dx}{d\ln(Q^2)}&
=&-B_0^d\,x^2-B_1^d\,x^3-B_{1,\lambda}^d x^2\lambda+\cdots\\
\nonumber
\beta^d_{\lambda}(x,\lambda)\equiv \frac{d\lambda}{d\ln(Q^2)}&
=&C_{\lambda}^d\,x\,\lambda+C_{\lambda\lambda}^d\,\lambda^2+\cdots
\end{eqnarray}
with
\begin{eqnarray}
\label{dual_coef}
B_0^d&=&\frac14\left(2N_f-3N_c\right)\\ \nonumber
B_1^d&=&\frac18\left(N_f-N_c\right)\left(2N_f-3N_c\right)-\frac{N_f}{8}\,
\frac{\left(N_f-N_c\right)^2-1}{N_f-N_c} \\ \nonumber
B_{1,\lambda}^d&=&\frac14N_f^2 \\ \nonumber
C_{\lambda}^d&=&-\frac12\frac{\left(N_f-N_c\right)^2-1}{N_f-N_c} \\ \nonumber
C_{\lambda\lambda}^d&=&\frac12\left(3N_f-N_c\right)
\end{eqnarray}
where $B_0^d$ and $B_1^d$ can be obtained by substituting
$N_c\longrightarrow N_f-N_c$ in $B_0$ and $B_1$, and the other
coefficients where calculated in \cite{KSV}. Note that in
(\ref{dual_coef}) we use $N_c$ to denote the number of colors in the
original (electric) theory, and thus the dual theory has an $SU(N_f-N_c)$
gauge symmetry. This is contrary to the notation used in \cite{KSV} that
corresponds to an $SU(N_c)$ gauge group in the magnetic theory. In
addition, note that in \cite{KSV} there is a typo in eq. (64), 
where a factor of two is missing in the second term in the second
equation\footnote{The authors thank D. Anselmi and R. Oehme for their
help on this matter.}. The correct
factor can be easily obtained by using eqs. (55), (62) and (63)
there. Our coefficients do agree with those in
\cite{Oehme_reduction}.

In order to study the analyticity structure of the coupling in the
magnetic theory one should, in principle, integrate the coupled
renormalization-group equation (\ref{coupled_RG}). This is, however,
rather complicated, and so we choose a simpler approach (which remains
to be further justified) based on the notion of Reduction of Couplings. 
  
It was recently shown by Oehme \cite{Oehme_reduction} that there is a
unique reduction of the coupled renormalization-group equation
(\ref{coupled_RG}) to a single-coupling equation such that 
the superpotential does not vanish, which is essential for duality.
Ref. \cite{Oehme_reduction} describes in detail how to apply the
general method of Reduction of Couplings to this problem. We shall use
here only the leading order relation between $\lambda$ and $x$.
To obtain the relation between the couplings one assumes
\beq
\lambda(x)= f(N_c,N_f)\,x +{\cal O}(x^2)
\label{reduction_form}
\eeq
and imposes the consistency condition,
\beq
\beta^d_{\lambda}\left(x,\lambda(x)\right)\,
=\,\frac{d\lambda(x)}{dx}\,\beta^d_x\left(x,\lambda(x)\right).
\label{reduction_condition}
\eeq  
Using (\ref{coupled_RG}), the condition (\ref{reduction_condition})
leads, at leading order, to: 
\beq
f(N_c,N_f) \left[C_{\lambda\lambda}^d
\,f(N_c,N_f)+C_{\lambda}^d +B_0^d\right]\,=\, 0
\label{red_con}
\eeq
and for a non-vanishing superpotential, the results is
\beq
f(N_c,N_f)\,=\,\frac{-B_0^d-C_{\lambda}^d}{C_{\lambda\lambda}^d}\,=\,
\frac{N_cN_f-N_c^2-2}{2\left(N_f-N_c\right)\left(3N_f-N_c\right)}.
\label{reduction_result}
\eeq
Note that $f(N_c,N_f)$ is positive in the entire conformal window.

With the result (\ref{reduction_result}) at hand we can substitute the
$\lambda$ term for $f(N_c,N_f)\,x$ in the equation of 
$\beta^d_x(x,\lambda)$ and obtain a
single-coupling renormalization-group equation which is valid up to
2-loop order:
\beq
\label{reduced_RG}
\beta^d_x(x,\lambda(x))\equiv \frac{dx}{d\ln(Q^2)}
=-B_0^d\,x^2-\tilde{B}_1^d\,x^3+\cdots
\eeq 
where 
\beq
\tilde{B}_1^d=B_1^d+B_{1,\lambda}^d\,f(N_c,N_f).
\label{B1_dual_reduced}
\eeq

Next, we would like to analyze the analyticity structure of the
coupling in the dual theory, using the reduced $\beta$ function
(\ref{reduced_RG}). 
Let us calculate first the condition for the dual 2-loop coupling to have a
causal analyticity structure (the analog of (\ref{R_SQCD}) in the
original theory).
The causality condition, $\tilde{B}_1^d/B_0^d<-B_0^d$, yields
\beq
18R^3-64R^2+\left[65+\frac{2}{N_c^2}\right]R-15<0  
\label{R_SQCD_d_eq}
\eeq
which again leads to an approximately $N_c$ independent 
critical value for $R$ for any possible value of $N_c$ 
(since $N_c^2\gg 2/65$), 
namely the 2-loop coupling in the dual theory is causal as long as
\beq
R\lsim 1.8357.
\label{R_SQCD_d}
\eeq
As with the original theory, the 2-loop causality region of the dual
perturbative coupling does not cover
the far-end of the window. Note (fig.~\ref{conformal_window}) 
that the regions of a causal 2-loop coupling
in the two dual descriptions, (\ref{R_SQCD}) and (\ref{R_SQCD_d}) do not
overlap. This fits the intuition on which duality is based, i.e. that when
one theory is weakly coupled its dual is necessarily strongly coupled.
Since we assume that within the window a consistent perturbation theory 
implies small non-perturbative effects, an overlap would lead to
contradiction: it would suggest that two different weakly coupled theories
can describe the same infrared physics.

In fig.~\ref{conformal_window} the 2-loop causality boundaries in the two
theories are very close. However, if one adopts a 
conservative attitude that perturbation theory actually breaks down 
above the 2-loop
causality boundary (taking into account the large 4-loop
correction) the perturbative regions of the two theories will be 
more separated.

One can also find the condition to have no space-like
Landau singularity in the 2-loop reduced coupling. The requirement
$\tilde{B}_1^d<0$ translates into the condition
\beq
3R^3-12R^2 +\left[13+\frac{1}{N_c^2}\right]R-3<0,
\label{R_space_like_SQCD_d_eq}
\eeq
which yields (for $N_c^2\gg 1/13$),
\beq
R\lsim 2.314.
\label{R_space_like_SQCD_d}
\eeq
Note that this line is below the top of the window, and thus in the
upper part of the 
conformal window the dual coupling has a space-like singularity. 

\subsection{Summary} 

To conclude this part, let us summarize the differences between QCD and SQCD
with respect to the analyticity structure of the coupling in comparison with 
the boundaries of the conformal window (fig.~\ref{conformal_window}). 

In QCD, the region of a causal 2-loop coupling covers the entire conformal 
window (supposing the lower boundary is determined by superconvergence:
$r=3.25$). As $r$ is reduced further (below $r\simeq3.17$), 
the 2-loop coupling develops a couple of Landau branch points at 
complex $Q^2$ values. At even lower $r$, below \hbox{$r\simeq 2.62$},
a Landau branch point appears on the space-like axis. 

Studying higher loop effects we showed that the 3-loop term is
important in the lower part of the conformal window, and so the 3-loop
coupling should be referred to as a zeroth order approximation in the
infrared. The next observation is that the 3-loop
coefficient is negative in the conformal window both 
in ${\hbox{$\overline{\hbox{MS}}\,$}}$ and in all the physical
effective charges for which the 3-loop coefficient has been computed.
This means that the 3-loop coupling is causal at least where the
2-loop coupling is, i.e. in the entire conformal window,
and in many cases, depending on $\beta_2$, 
also somewhat below this region into the upper part of the confining phase. 
The 3-loop solution is reliable according to the usual perturbative
justification: the 4-loop term in the
$\beta$ function, at least in ${\hbox{$\overline{\hbox{MS}}\,$}}$, 
is small enough not to affect the 3-loop solution.  

In SQCD, the region of a causal 2-loop coupling $1.8377\lsim R<3$
does not cover the lower part of the conformal window (the lower
boundary is at $R=1.5$).  
Below $R\simeq 1.8377$ the 2-loop coupling develops a couple of Landau branch 
points at complex $Q^2$, and below $R\simeq 1.5$, i.e. below the
conformal window (see eq. (\ref{r_cd})) the 2-loop coupling has a space-like 
Landau singularity. Studying higher orders we find that the 3-loop
term is significant, and like in QCD it leads to a smaller
coupling and to a larger causality region. But since the value of the
coupling is still not small enough, and the 4-loop term is large, the
3-loop solution cannot be trusted. This means that the 
perturbative analysis in the electric theory 
is reliable only in the upper part of the conformal window. 
In the dual (magnetic) theory the reduced 2-loop coupling is causal
only in the region: $1.5 \lsim R\lsim 1.8357$.
This coupling even has a space-like Landau singularity inside the
window, for $R\gsim 2.314$.

Our main conclusions from this analysis are the following:
\begin{description}
\item{(a) } In QCD perturbation theory seems consistent in the infrared 
within the entire conformal window, and even somewhat below it.
It then seems natural to assume that non-perturbative corrections 
are small, at least within the conformal window. 
\item{(b) } The previous assumption implies that in QCD the fields
are, in some sense, weakly coupled even at the bottom of the window. 
This is contrary to SQCD where the electric fields are strongly 
coupled at the bottom of the window, one of the assumptions on which 
duality is based (see (d) below).
We conclude that in QCD there is no dual description of the infrared in 
terms of some alternative degrees of freedom which are weakly coupled 
near the bottom of the window. 
\item{(c) } We found that the fixed-point in SQCD at the far-end of
the conformal window cannot be explained in terms of the perturbative $\beta$
function.  
\item{(d) } The regions where the electric and magnetic 2-loop
couplings in SQCD are causal do not overlap. 
Perturbation theory is never meaningful in the infrared in both the 
electric and magnetic descriptions of the same model. 
This is in accordance with the assumption on which duality is based that 
when the electric theory is weekly coupled, the magnetic is
necessarily strongly coupled and vice-versa.
\item{(e) } In SQCD perturbation theory signals its own inapplicability 
indicating that the coupling becomes strong within the window. This
fits the same general philosophy on which the assumption in (a) is based:
the strong coupling nature of the theory at the bottom of the conformal 
window should manifest itself already in perturbation theory.
\end{description}

\section{Banks-Zaks expansion in SQCD vs. QCD}

In the previous sections we saw that in QCD perturbation theory 
yields a consistent description of the infrared physics
even in the lower part of the conformal window: the coupling
is causal and stable with respect to higher-loop corrections.
On the other hand, in SQCD causality cannot be achieved at
the perturbative level in the lower part of the conformal window.

In order to examine the effect of higher order corrections 
we used an explicit solution of the equation
$\beta(x)=0$ in the ${\hbox{$\overline{\hbox{MS}}\,$}}$ scheme and in
physical schemes in QCD, and in the DRED scheme in SQCD.
Another natural way to study the value of the physical quantities
in the infrared is the Banks-Zaks
expansion, i.e. a power series solution to the equation $\beta(x)=0$,
in terms of the distance from the top of the conformal window. In QCD,
the expansion parameter is \hbox{$\epsilon\equiv(11/2)-(N_f/N_c)=6\beta_0/N_c$}  and the expansion has the form:
\beq
x_{\FP}\, =\, z_1\epsilon\,+\,z_2\epsilon^2\,
+\,z_3\epsilon^3\,+\cdots
\label{BZ}
\eeq 
where $z_i$ are independent of $N_f$.
Since the coefficients of the $\beta$ function are polynomials
in $N_f$, it is possible to write them as follows. The 2-loop coefficient: 
\beq
c=\frac{\beta_1}{\beta_0}=-\frac{1}{a_0}+c_{1,0}
\label{c_a0}
\eeq
where $a_0$ is proportional to $\epsilon$ (and to $\beta_0$) and
$c_{1,0}$ is independent of $N_f$. The 3-loop coefficient:
\beq
c_2=\frac{\beta_2}{\beta_0}
=c_{2,-1}\frac1{a_0}+c_{2,0}+c_{2,1}a_0+c_{2,2}a_0^2,
\label{c_2_a0}
\eeq
where $c_{2,i}$ are independent on $N_f$, and so on.
Then the leading terms in the Banks-Zaks 
expansion for a generic effective charge are \cite{BZ_grunberg,CaSt},  
\beq
x_{\FP}\,=\,a_0\,+\,\left(c_{1,0}+c_{2,-1}\right)a_0^2\,+\, \cdots 
\label{2_leading}
\eeq
We identify $a_0=z_1\epsilon$ and note that $z_1$ is the 
same for any effective-charge (or coupling) due to the universality of $c$. 
However, already $z_2$ depends on the effective-charge (or
coupling) under consideration -- according to eq. (\ref{2_leading}) it 
depends on the 3-loop coefficient of the effective-charge $\beta$ function.

We stress that the ultimate justification of the presence of a
fixed-point near the top of the conformal window, 
and thus of the very existence of the conformal window, 
is through this expansion \cite{BZ,BZ_grunberg}.  
On the other hand, it is a priori not at all clear
how far into the conformal window one can trust the expansion. 
We will be interested in particular in calculating the coupling and
the critical exponent in QCD at the bottom of the conformal window and
in estimating the reliability of this calculation. We
will show that a calculation of this sort cannot be done
in SQCD in the lower part of the conformal window.

\subsection{Banks-Zaks expansion for the coupling}

\subsubsection{Banks-Zaks expansion for the coupling in QCD}

As in the previous sections we start by considering 
the ${\hbox{$\overline{\hbox{MS}}\,$}}$ scheme. 
The advantage is that the coefficients of the $\beta$ function are known up to 
4-loop order \cite{fourloops}. This will enable us to compare
the infrared limit obtained from the explicit solution of
$\beta(x)=0$ (fig.~\ref{QCD_IR_epsilon}) which seems quite reliable at
the 3-loop and 4-loop orders, to that of the Banks-Zaks partial-sums.
A disadvantage of this scheme is that the coupling constant
is not directly related to any measurable quantity. 
The dependence of the Banks-Zaks expansion on the effective charge or
coupling under consideration, which was investigated in \cite{FP}, 
first appears at the next-to-leading order term in the expansion -- 
see eq. (\ref{2_leading}). This dependence becomes significant at the
next-to-next-to-leading order level. 

According to \cite{FP} the next-to-next-to-leading order coefficient 
in the Banks-Zaks expansion for the ${\hbox{$\overline{\hbox{MS}}\,$}}$ 
coupling is rather large, making the corresponding term in the expansion
comparable to the leading order terms already within the conformal
window. Here we shall further analyze the expansion for the
${\hbox{$\overline{\hbox{MS}}\,$}}$ coupling explaining the source of
the large next-to-next-to-leading coefficient.
For physical effective charges this coefficient is smaller 
than in ${\hbox{$\overline{\hbox{MS}}\,$}}$, hence the expansion is 
more reliable.

The coefficients of the $\beta$ function in the
${\hbox{$\overline{\hbox{MS}}\,$}}$ scheme are known up to 4-loop
order \cite{threeloops,fourloops}.
The three first Banks-Zaks coefficients
in the expansion of $x^{\FP}_{\MSbar}$ (\ref{BZ}) are then determined:
\begin{eqnarray}
\label{BZ_QCD}
z_{1}& =&  \frac{16}{3} \,\frac{N_c}{25\,{N_c}^{2} -
11}\\ \nonumber
z_{2}& =& \frac{16}{27} \,\frac{N_c \,\left[548\,{N_c}^{4} - 1066\,{N_c}^{2}
 + 231\right]}{\left(25\,{N_c}^{2} - 11\right)^{3}}\\ \nonumber
z_{3}& =&  \frac{32}{243}\,\frac{N_c\, J}{\left(25\,{N_c}^{2} - 11\right)^{5}}
\end{eqnarray}
with 
\begin{eqnarray*}
\lefteqn{J\,=\,52272+(389235-1341648\zeta_3)N_c^2+
(-719758+3362832\zeta_3)N_c^4}\\ \nonumber
&&+(-1148400\zeta_3-1105385)N_c^6
+(990000\zeta_3+730529)N_c^8
\end{eqnarray*}

Let us examine whether the Banks-Zaks expansion (\ref{BZ}) is still
reliable at the bottom of the conformal window. 
Table 1 summarizes the results for $N_cx_{\footnotesize{\MSbar}}^{\FP}$
(this normalization is used in order to consider both finite $N_c$
cases and the large $N_c$ limit) 
according to (\ref{BZ}) and (\ref{BZ_QCD}) at the lower boundary 
of the conformal window, namely at $\epsilon=11/2-13/4=9/4$ for 
$N_c=2$, $N_c=3$ and 
$N_c\longrightarrow\infty$. The results are presented as a function of
order in $\epsilon$: order $\epsilon$ stands for the leading term
in (\ref{BZ}), order $\epsilon^2$ stands for the sum of the first two
terms in (\ref{BZ}), an so on.
\begin{table}[H]
\[
\begin{array}{|c|c|c|c|}
\hline
\mbox{order}&N_c=2&N_c=3&N_c\longrightarrow\infty\\
\hline
\epsilon&0.539&0.505&0.480 \\
\hline
\epsilon^2&0.620&0.601&0.585 \\
\hline
\epsilon^3&1.03&0.933&0.880 \\
\hline
\end{array}
\]
\caption{$N_c x^{\FP}_{\MSbar}$ in QCD at the bottom of the
  conformal window as a function of order in the Banks-Zaks expansion. } 
\end{table}

Our first conclusion from table 1 is that there is no significant 
dependence on $N_c$: there is no much difference between 
$N_c x^{\FP}_{\MSbar}$ for $N_c=2$ and for $N_c=\infty$.  

As mentioned above, the ${\cal O}(\epsilon^3)$ term at the
bottom of the window is larger than the ${\cal O}(\epsilon^2)$ term
there. Note that it is also comparable to the leading 
${\cal O}(\epsilon)$ term. This clearly raises doubts concerning the
reliability of the expansion. On the other hand, solving explicitly
$\beta(x)=0$ we found in Sec. 2 that the 4-loop fixed-point value is
almost identical to the 3-loop one down to the bottom of the conformal
window (fig.~\ref{QCD_IR_epsilon}). This calls for a more detailed
examination of the relation between the Banks-Zaks expansion and the
explicit solution, which we conduct in the next section.  

\subsubsection{The reliability of the Banks-Zaks expansion in QCD at
the bottom of the conformal window}

The purpose of this section is to understand the reason for
the large ${\cal O}(\epsilon^3)$ term in the Banks-Zaks expansion in 
${\hbox{$\overline{\hbox{MS}}\,$}}$, and finally to estimate the 
reliability of the fixed-point value.
The analysis we present is for the case $N_c \longrightarrow \infty$, 
but the results for low $N_c$ are qualitatively the same.

Let us compare first the numerical values obtained at the bottom of the
window from the explicit solution vs. the corresponding
partial sum in the Banks-Zaks expansion:
\begin{table}[H]
\[
\begin{array}{|c|c||c|c|}
\hline
\mbox{order}& &\beta(x)=0 &\\
\hline
\epsilon&0.480& \mbox{2-loop} &2.1818\\
\hline
\epsilon^2&0.585&\mbox{3-loop}& 0.7495\\
\hline
\epsilon^3&0.880&\mbox{4-loop}& 0.7667\\
\hline
\end{array}
\]
\caption{$N_c x^{\FP}_{\MSbar}$ in large $N_c$ QCD at the bottom of the
  conformal window as a function of order in the Banks-Zaks expansion
  and from an explicit solution of the equations $\beta(x)=0$ for the
  truncated $\beta$ function at each order.} 
\end{table}
This comparison is shown also in fig.~\ref{QCD_convergence_BZ_loops}.
We see that the two calculation procedures agree. Referring to the
explicit solution as the best estimate at hand, we can estimate the
uncertainly in the value of the infrared coupling from the difference
between the two calculation procedures. 
For the $x^{\FP}_{\MSbar}$ the uncertainty is no more than $\pm 25\%$.        

Let now investigate the relation between the explicit solutions 
and the Banks-Zaks expansion. At 2-loop order, 
the functional form of the fixed-point value in the large $N_c$ limit is 
\beq
N_cx(0)=N_c \left(\frac{-1}{c}\right)=\frac{16\epsilon}{75-26\epsilon}.
\label{2_loop_fp_epsilon}
\eeq 
At higher loop orders, the result is a more complicated function of
$\epsilon$. At any order the explicit solution has a finite
convergence radius in powers of $\epsilon$, and thus we expand it, and
compare the expansion to the function itself. Such a comparison is shown
in fig.~\ref{QCD_convergence_BZ_order} at the bottom of the conformal
window, i.e. for $\epsilon=2.25$. 

In the upper plot, corresponding to the 2-loop case, we see that the 
expansion in $\epsilon$ converges very slowly to the explicit
solution. 
This can be understood knowing that $N_c x(0)$ is a geometrical series
in $\epsilon$ (\ref{2_loop_fp_epsilon}) and that
$\epsilon$ at the bottom of the window is already quite close to the
convergence radius which is $\epsilon=75/26\simeq 2.88$, the point where $c$
vanishes. Since we know from the comparison with the explicit
solutions at higher orders that close to the bottom of the conformal
window the 2-loop value for $N_cx(0)$ is unrealistically
large\footnote{This is related to the discussion in Sec. 2 concerning 
the necessity to start from the 3-loop term in order 
to establish perturbative causality in the lower part of the window.} 
we should not regard the slow convergence
of the series in $\epsilon$ corresponding to (\ref{2_loop_fp_epsilon})
as indicative of a problem of the Banks-Zaks series as a whole. It just
means that higher orders are important.  

In the 3-loop case in fig.~\ref{QCD_convergence_BZ_order} 
(middle plot) the Banks-Zaks partial sum at order ${\cal
O}(\epsilon^2)$ is much closer to the explicit solution and the
convergence at higher orders in $\epsilon$ is much accelerated as
compared to the 2-loop case. 

In the 4-loop case in fig.~\ref{QCD_convergence_BZ_order}
(lower plot) the partial sums of the $\epsilon$ expansion diverge
badly beyond the  ${\cal O}(\epsilon^3)$ term or so.  
The reason is that the
convergence radius of the $\epsilon$ series of the explicit solution
is about $\epsilon\simeq 1$, i.e. significantly smaller than 
$\epsilon=2.25$ which corresponds to the bottom of the window and to 
fig.~\ref{QCD_convergence_BZ_order}.  
This also explains why the
${\cal O}(\epsilon^3)$ term in the Banks-Zaks series, which is fully
determined at the 4-loop level, 
is larger than the ${\cal O}(\epsilon^2)$ term. 
The explicit solution is a well defined function of $\epsilon$
in the entire conformal window in all the cases considered. It turns
out however that in the 4-loop case this function does not have a 
converging power expansion beyond $\epsilon\simeq 1$. 
This fact is  
shown also in fig.~\ref{QCD_IR_epsilon}: around $\epsilon\simeq 1$ the series
departs from the explicit solution itself. 

We note that for the available examples the $\epsilon$ series that
correspond to increasing loop-order solutions have an ever decreasing 
convergence radii: it is $\epsilon\simeq 2.88$ in the 2-loop case,
$\epsilon \simeq 2.787$ in the 3-loop case and $\epsilon\simeq 1$ in
the 4-loop case.
This may be related to large order behavior of series:
since the Banks-Zaks expansion is based on the factorially 
growing perturbative coefficients, it is natural to expect that 
it is also an asymptotic series with {\em zero radius of convergence}. 
Such a behavior will be avoided only if some systematic cancellation of the 
factorially growing ingredients occurs.  If indeed the asymptotic
nature of the Banks-Zaks series is reached at the order ${\cal O}(\epsilon^3)$
the best estimate of the fixed-point value from the expansion is obtained
by truncating the series after the minimal term, in this case, the
next-to-leading term. 

A comparison between the fixed-point value from the Banks-Zaks
expansion and the explicit solution of $\beta(x)=0$ can be also
conducted in physical renormalization schemes. In the
absence of full 4-loop perturbative coefficients, one cannot obtain 
an explicit solution at the 4-loop level. On the other hand, the 
${\cal O}(\epsilon^3)$ is calculable \cite{BZ_grunberg,CaSt,FP} and
thus the next-to-next-to-leading order partial sum 
can be compared with the explicit solution of the 3-loop effective
charge $\beta$ function.
Such a comparison was performed in \cite{FP} for the effective charge
which is defined from the vacuum-polarization D-function. As shown in fig.~7
there, the two calculation methods nicely agree down to the bottom of 
the conformal window ($N_f\simeq 10$ in the figure) and even below.   

As noted above, in physical renormalization schemes the Banks-Zaks 
coefficients (and in particular the next-to-next-to-leading coefficients) 
are smaller than in  ${\hbox{$\overline{\hbox{MS}}\,$}}$
\cite{FP,CaSt}, and so the expansion seems more reliable.  
For example, for $N_c=3$ we have \cite{FP}:
\begin{eqnarray}
\label{BZ_physical_schemes}
x_{\FP}^{\MSbar}\,&=&\,a_0\,+\,1.14 \,a_0^2\,+\,23.27 \,a_0^3\,+\, 
\cdots \\ \nonumber 
x_{\FP}^{D}\,&=&\,a_0\,+\,1.22 \,a_0^2\,+\, 0.23 \,a_0^3\,
+\, \cdots \\ \nonumber
x_{\FP}^{V}\,&=&\,a_0\,-\,0.86 \,a_0^2\,+\,10.99 \,a_0^3\,
+\, \cdots 
\end{eqnarray}
where D stand for the effective charge defined from the vacuum
polarization D-function and V stands for the one defined from the heavy
quark potential. In fact, the ${\cal O}(a_0^3)$ coefficient in $x_{\FP}^{V}$
is the largest amongst all the ${\cal O}(a_0^3)$ coefficients for the
effective charges considered in \cite{FP}\footnote{The result
presented above for the ${\cal O}(a_0^3)$ coefficient in $x_{\FP}^{V}$
is different from the one in \cite{FP}. The latter was calculated
based on a wrong 2-loop coefficient, which has now been corrected
thanks to \cite{V-scheme-corrected}.}.

We conclude that calculation of infrared quantities can be performed either
as an explicit solution of the equation $\beta(x)=0$ 
or by the Banks-Zaks expansion. Although
the expansion probably has a zero convergence radius in general, and
bad convergence properties already for the available 4-loop example
(${\hbox{$\overline{\hbox{MS}}\,$}}$), it seems to give a reasonable 
estimate at the next-to-leading and the next-to-next-to-leading orders
within the entire conformal window. 
Infrared quantities appear to be perturbatively calculable in general 
even at the bottom of the conformal window. Note, however, that 
the accuracy is observable dependent. Some quantities, like 
the vacuum polarization D-function, can be determined with high
accuracy, whereas for others the accuracy is not as good: as mentioned
above, the ${\hbox{$\overline{\hbox{MS}}\,$}}$ coupling can be determined
within $\pm 25 \%$ accuracy.

\subsubsection{Banks-Zaks expansion for the coupling in SQCD}

Let us now turn to the supersymmetric case and 
consider the Banks-Zaks expansion for the value of the DRED
coupling at the fixed-point. The expansion parameter is 
$\delta \equiv 3-R = 3-(N_f/N_c)$:  
\beq
x_{\DRED}^{\FP}\, =\, Z_1\delta\,+\,Z_2\delta^2\,
+\,Z_3\delta^3\,+\,{\cal{O}}(\delta^4).
\label{BZ_SQCD_ser}
\eeq
The coefficients of the $\beta$ function up to 4-loop are taken 
from \cite{DRED}. The resulting Banks-Zaks coefficients read:
\begin{eqnarray}
\label{BZ_SQCD}
Z_{1}& =& \frac{2}{3} \, \frac{N_c}{N_c^2-1} \\ \nonumber
Z_{2}& =& \frac{1}{3}\, \frac{N_c}{\left(N_c^2-1\right)} \\ \nonumber
Z_{3}& =& \frac{1}{54} \,\frac{N_c\, \left[(17 +18\zeta_3)N_c^4
+(-25 +18 \zeta_3) N_c^2 +8\right]}{\left(N_c^2-1\right)^3}\\ \nonumber
\end{eqnarray}

Table 3 summarizes the results
for $N_cx_{\DRED}^{\FP}$,
 according to (\ref{BZ_SQCD_ser}) and (\ref{BZ_SQCD}), 
at the bottom of the conformal window, i.e. at $\delta=3-3/2=3/2$: 
\begin{table}[H]
\[
\begin{array}{|c|c|c|c|}
\hline
\mbox{order}&N_c=2&N_c=3&N_c\longrightarrow\infty\\
\hline
\delta&1.33&1.13&1 \\
\hline
\delta^2&2.33&1.97&1.75 \\
\hline
\delta^3&8.00&5.38&4.16 \\
\hline
\end{array}
\]
\caption{$N_c x^{\FP}_{\DRED}$ in SQCD at the bottom of the
  conformal window as a function of order in the Banks-Zaks expansion. } 
\end{table}

There is a clear contrast between the supersymmetric case of table 3 and the
non-supersymmetric case of table 1.
Table 3 shows that the Banks-Zaks series for 
$N_c x^{\FP}_{\DRED}$ 
at the bottom of the conformal window cannot be trusted at all,
since the next-to-leading term is comparable to the leading one and the
third order term is much larger than both.
In addition, the value of the coupling itself (as much as it can be
determined) is larger than in QCD.

It is interesting to compare between the explicit solutions to the
equations $\beta(x)=0$ at increasing loop order
(fig.~\ref{SQCD_IR_delta}), and the Banks-Zaks expansion.
In the following table we show the values of the infrared coupling
at the bottom of the window as determined by the two methods: 
\begin{table}[H]
\[
\begin{array}{|c|c||c|c|}
\hline
\mbox{order}&&\beta(x)=0&\\
\hline
\delta&1&\mbox{2-loop}&\infty \\
\hline
\delta^2&1.75&\mbox{3-loop}& 4\\
\hline
\delta^3&4.16&\mbox{4-loop}& \mbox{no solution}\\
\hline
\end{array}
\]
\caption{$N_c x^{\FP}_{\DRED}$ in SQCD for $N_c \longrightarrow \infty$
  at the bottom of the
  conformal window as a function of order in the Banks-Zaks expansion,
  and from the explicit solution of $\beta(x)=0$.} 
\end{table}

It is clear from this table and from fig.~\ref{SQCD_IR_delta}
that the perturbative analysis fails to determine 
the infrared value of the coupling in the lower part of the conformal
window. It thus seems, also from this point of view, that perturbation theory 
is inapplicable to describe the infrared physics there.

\subsubsection{Banks-Zaks expansion for the coupling in the magnetic
theory (dual SQCD)}

In a similar manner we consider the Banks-Zaks series in the dual
theory, where the expansion parameter is 
$\delta_d=R-(3/2)$\footnote{Note that both expansion parameters
$\delta$ and $\delta_d$ are chosen to be positive inside 
the conformal window.},
\beq
x^{\FP}_{\rm {dual}}\, 
=\, Z_1^d\delta_d\,+\,Z_2^d\delta_d^2\,
+\,Z_3^d\delta_d^3\,+\,{\cal{O}}(\delta_d^4).
\label{BZ_SQCD_ser_dual}
\eeq
The coefficients can be calculated either from the reduced $\beta$
function (\ref{reduced_RG}), or directly from the coupled
$\beta$ function (\ref{coupled_RG}), assuming both
infrared couplings are vanishingly small.
Since the $\beta$ function in the magnetic theory is
known at present only up to the next-to-leading order term, only the leading
order coefficient in the Banks-Zaks expansion can be calculated.
The result is:
\beq
Z_1^d=\frac{112}{3}\,\frac{N_c}{N_c^2-4}
\label{BZ_SQCD_dual}
\eeq
The infrared value of the Yukawa coupling is given by 
\beq
\lambda^{\FP}\,=\,\frac{16}{3}\frac{1}{N_c}\,\delta_d+{\cal O}(\delta_d^2).
\label{lambda_FP}
\eeq

Let us now examine the magnitude of the infrared coupling in the magnetic
theory at the top of the conformal window 
(having only one term, we cannot investigate the behavior of the
series as we did for the electric theory and for the
non-supersymmetric case).
Using the leading term in (\ref{BZ_SQCD_ser_dual}) with
(\ref{BZ_SQCD_dual}) and \hbox{$\delta_d=3/2$} we find that
for  \hbox{$N_c=3$},
\hbox{$x_{\FP}\simeq168/5=33.6$}, and for 
\hbox{$N_c=\infty$}, \hbox{$N_c x_{\FP}\simeq 56$}. 
In both cases, it is clear that the coupling is much too large to be
perturbative (which also implies that these values are meaningless).
The conclusion is that the fixed-point of the dual theory cannot be 
described by perturbation theory at the far-end of the window.

An interesting unrelated observation is that for $N_c=2$, the
Banks-Zaks expansion is completely ill-defined due to the pole at $N_c^2=4$ in
(\ref{BZ_SQCD_dual}). In the absence of the Banks-Zaks expansion it seems
hard to establish the existence of a fixed-point. In fact, as we
explain below, the problem is specific to the point around 
which the expansion is done, and therefore it may not imply anything special 
for the rest of the conformal window for $N_c=2$\footnote{The authors
are in debt to D. Anselmi for explaining this point.}. The
original theory in this case (at the bottom of the conformal window) 
is an $SU(2)$ gauge theory with
$N_f=3$. The implied dual theory has a color group of $N_f-N_c=1$, which means
that there are no gluons. Mathematically, this appears as an
ill-defined expansion since the point where the next-to-leading 
coefficient of the $\beta$ function $\tilde{B}_1^d$ vanishes
(see eq. (\ref{B1_dual_reduced})) coincides with the point where
$B_0^d$ vanishes\footnote{For any $N_c>2$, 
$\tilde{B}_1^d$ becomes negative already at lower $N_f/N_c$, before
$B_0^d$ vanishes.}, and thus the ratio $B_0^d/\tilde{B}_1^d$ which is
usually used to
define the expansion parameter $\delta_d$ is not arbitrarily 
small near the point $B_0^d=0$ but is finite there. 

\subsection{Banks-Zaks expansion for the critical exponent}

The critical exponent $\gamma$ has a
special status since it is a {\em universal quantity} \cite{Gross}: 
it determines the
rate at which {\em any} perturbative coupling or effective-charge
approaches its infrared limit\footnote{$\gamma$ in QCD was
  discussed in various papers; see for instance 
\cite{BZ_grunberg,CaSt,Chyla,FP}.}.
 In addition, discussing the analyticity structure of the coupling
we found that the value of $\gamma$ is indicative of a
causal coupling. 
Thus it is interesting to study the Banks-Zaks expansion and its
break-down for this particular quantity.

\setcounter{footnote}{0} 
Let us start with a brief review of the definition and the basic
properties of $\gamma$\footnote{The notation is again that of QCD, but
the same equations are relevant in SQCD, with the replacement 
of $\epsilon$ by $\delta$, $\beta_i$ by $B_i$, $c$ by $C_1$, and so on.}.
The critical exponent is defined as the derivative of 
the $\beta$ function,
\beq
\beta(x)=-\beta_0x^2\left(1+cx+c_2x^2+\cdots\right)
\eeq
at the fixed-point:
\beq
\gamma\equiv\left.\frac {d\beta(x)}{dx}\right\vert_{x=x_{\FP}}=
-\beta_0 x_{\FP}\left[2+3\,c\,x_{\FP}+4\,c_2\,\left(x_{\FP}\right)^2
+\cdots \right]
\label{gamma_def}
\eeq
from which eq. (\ref{first_x_x_fp}) follows.

As already mentioned $\gamma$ is universal, i.e. 
independent of the renormalization scheme. 
To be precise, this statement is true so long as the
transformations relating the different schemes are non-singular (see
ref. \cite{Gross,Chyla} and appendix B in ref. \cite{CaSt} and references
therein).

The Banks-Zaks expansion for $\gamma$ can be calculated using  
(\ref{gamma_def}) together with the Banks-Zaks series for
$x_{\FP}$, yielding a Banks-Zaks series of the form,
\beq
\gamma=g_1\epsilon^2+g_2\epsilon^3+g_3\epsilon^4+\cdots
\label{gamma_BZ}
\eeq
Note that contrary to a generic effective charge, 
the expansion for $\gamma$ begins with an $\epsilon^2$ term. A further
difference is that the coefficients of (\ref{gamma_BZ}) have an
additional factor of $N_c$, as compared to those of (\ref{BZ}).

It was shown in \cite{BZ_grunberg} 
that the coefficients $g_i$ are universal, i.e. they are the same for
any effective-charge $x$. This is in agreement with what is expected
on general grounds, since $\gamma$ itself is independent of the
renormalization scheme in which the $\beta$ function is defined, and
the expansion parameter $\epsilon$ is a well defined physical quantity.

An additional interesting observation \cite{BZ_grunberg} is that the
first two terms in the Banks-Zaks expansion for $\gamma$ are
determined from the 2-loop $\beta$ function:  
\beq  
\gamma\,=\,g_1\epsilon^2+g_2\epsilon^3+\cdots\,=
\,(g_1/z_1^2)\,\left[a_0^2\,+\,c_{1,0}a_0^3\,+\,\cdots\right]
\label{2_leading_gamma}
\eeq
where $a_0=z_1\epsilon$ and $c_{1,0}$ are defined in (\ref{c_a0}). 
Since $g_2$ is fixed by the 2-loop
$\beta$ function which is the leading order in which the Banks-Zaks
fixed-point can be discussed, it makes sense to regard the first two 
orders $g_1\epsilon^2+g_2\epsilon^3$ together as the leading term. 
We shall see below that in both QCD and SQCD $g_2\epsilon^3$ is
comparable to $g_1\epsilon^2$ for values of $\epsilon$ such that 
the expansion for the coupling is
still reliable\footnote{In QCD it is the case in all the physical
renormalization schemes that where examined in \cite{FP}, since always
$c_{2,-1}<0$. Thus it turns out that the next-to-leading coefficient in
(\ref{2_leading}) is smaller in absolute value than the one in 
(\ref{2_leading_gamma}).}. However, according to the explanation above
this should not be regarded as an indication of the break down of the 
series -- it is the magnitude of the next term $g_3\epsilon^4$, that
depends also on the 3-loop and 4-loop coefficients of the $\beta$ function, 
which must be examined in order to assess the reliability the expansion.

\subsubsection{The critical exponent in QCD}

Again, we start with QCD where the coefficients of the Banks-Zaks
series for $\gamma$ in (\ref{gamma_BZ}) 
are\footnote{The third order coefficient of the 
Banks-Zaks expansion for $\gamma$ has been calculated for the first
time in \cite{CaSt}.}:
\begin{eqnarray}
\label{BZ_QCD_gamma}
g_{1}& =&   \frac {8}{9} 
\, \frac{N_c^2}{25\,N_c^{2} - 11} 
 \\ \nonumber
g_{2}& =&  \frac {16}{27}\,
\frac {\,N_c^2\,(13\,N_c^{2} - 3)}{(25\,N_c^{2} - 11)^{2}}
\\ \nonumber
g_{3}& =-& \frac{8N_c^{2}\, H}{243\,(25\,N_c^{2} - 11)^{5}} 
\end{eqnarray}
with 
\begin{eqnarray*}
\lefteqn{H=3993 + (571516 - 894432\,\zeta_3)\,N_c^{2} + ( - 
1599316 + 2241888\,\zeta_3)\,N_c^{4} + } \\
 & & ( - 765600\,\zeta_3 + 865400)\,N_c^{6} + (660000\,
\zeta_3 - 366782)\,N_c^{8} 
\end{eqnarray*}

Our aim is to see whether $\gamma$ can be calculated from this
expansion even at the bottom of the conformal window, and then with
what accuracy. 
Fig.~\ref{combined_gamma_BZ} (upper plot) shows, in addition to the
results of the explicit calculation in various schemes, 
the following Banks-Zaks partial sums:
\hbox{$g_1\epsilon^2$}, \hbox{$g_1\epsilon^2+g_2\epsilon^3$}, and
\hbox{$g_1\epsilon^2+g_2\epsilon^3+g_3\epsilon^4$} as a function of
$\epsilon$. The next-to-leading term is
relatively large, but as explained in the previous section this should
not be taken as an indication for the breakdown of the expansion.
The relevant observation is that the 
next-to-next-to-leading term is just a small correction. At this level
the Banks-Zaks series for $\gamma$ seems reliable.

The comparison between the explicit calculation based on a truncated 
$\beta$ function and the Banks-Zaks partial sums, shown in 
fig.~\ref{combined_gamma_BZ} (upper plot) raises again the question of
the relation between the two calculation procedures, especially in the
${\hbox{$\overline{\hbox{MS}}\,$}}$ scheme.

The table below summarizes the numerical values obtained at the bottom of the
conformal window (like the plot, the numbers correspond to 
$N_c\longrightarrow \infty$ but the results at low $N_c$ are similar). 
\begin{table}[H]
\[
\begin{array}{|c|c||c|c|c|c|c|c|}
\hline
\mbox{order}& &\beta(x)=0 & ${\hbox{$\overline{\hbox{\small MS}}\,$}}$
&{\small D}&{\small Bj}&{\small F_1}&{\small V}       \\
\hline
\epsilon^2   &0.180   &               & & & & & \\
\hline
\epsilon^3 &0.320   &\mbox{2-loop} &0.818&0.818&0.818&0.818&0.818  \\
\hline
           &        &\mbox{3-loop} & 0.466&0.330&0.340&0.337&0.413 \\
\hline
\epsilon^4 &0.284   &\mbox{4-loop} & 0.463 &&&&\\
\hline
\end{array}
\]
\caption{$\gamma$ in large $N_c$ QCD at the bottom of the
  conformal window as a function of order in the Banks-Zaks expansion
  and as an explicit calculation from the truncated $\beta$
  function in ${\hbox{$\overline{\hbox{MS}}\,$}}$ and in various
  physical schemes: $D$ - vacuum polarization D-function, $Bj$ and
  $F_1$ - polarized and non-polarized Bjorken sum-rules and $V$ -
  heavy quark potential. The gaps in the table are due to the fact 
that the next-to-leading term ${\cal O}(\epsilon^3)$ of the Banks-Zaks
  series depends only on the 2-loop $\beta$ function, while the 
next-to-next-to-leading term ${\cal O}(\epsilon^4)$ is
 determined by the 4-loop $\beta$ function \cite{BZ_grunberg,CaSt,FP}. 
} 
\end{table}
Considered separarately, both the Banks-Zaks expansion and the explicit
calculation in ${\hbox{$\overline{\hbox{MS}}\,$}}$ seem reliable. 
Still the disagreement between them is about $40\%$. 
In order to understand better the source of this discrepancy we
compare in fig.~\ref{QCD_convergence_BZ_gamma_order} the explicit results
for $\gamma$ with the partial sums in the $\epsilon$ expansion of
these results, at the bottom of the window.
In the 2-loop (upper plot) and 3-loop (middle plot) cases the 
$\epsilon$ series converges to the value for $\gamma$, while in the
4-loop case, the series diverges since its convergence radius is
smaller than the value of $\epsilon$ at the bottom of the window, 
$\epsilon=2.25$.   

The comparison in fig.~\ref{QCD_convergence_BZ_gamma_order} 
suggests that the explicit calculation (right column in the table) 
is equivalent to some resummation of higher order terms in $\epsilon$,
and explains the disagreement between the two calculation procedures. 
Such a resummation is necessarily scheme dependent since it should reflect
the spread between the different schemes when using a truncated
$\beta$ function. Finally, at the available order in perturbation
theory we can determine the critical exponent to be $\gamma= 0.4\pm 0.1$.

\subsubsection{The critical exponent in SQCD}

In the SQCD case, in the original (electric) theory, the expansion for the
critical exponent is
\beq
\gamma=G_1\delta^2+G_2\delta^3+G_3\delta^4+\cdots
\label{gamma_SQCD_BZ}
\eeq
and the coefficients $G_i$  are:
\begin{eqnarray}
\label{BZ_SQCD_gamma}
G_{1}& =& \frac{1}{6} \, \frac{N_c^2}{\left(N_c^2-1\right)} \\ \nonumber
G_{2}& =& \frac{1}{18}\, \frac{N_c^2\left(2N_c^2-1\right)}
{\left(N_c^2-1\right)^2} 
\\ \nonumber
G_{3}& =& -\frac{1}{216} \,
\frac{N_c^2\, \left[(-1 +18\zeta_3)N_c^4+(2 +18 \zeta_3) N_c^2-5\right]}
{\left(N_c^2-1\right)^3}\\ \nonumber
\end{eqnarray}

The numerical values of $\gamma$ as calculated from the partial sums
\hbox{$G_1\delta^2$}, \hbox{$G_1\delta^2+G_2\delta^3$}, and  
\hbox{$G_1\delta^2+G_2\delta^3+G_3\delta^4$} 
is shown in fig.~\ref{combined_gamma_BZ} (lower plot) 
as a function of $\delta$ within the conformal window. It is clear
from the plot that the expansion is useless at the bottom of the
window since $G_3\delta^4$ is comparable to $G_1\delta^2$ and to $G_2\delta^3$.

\subsubsection{The critical exponent in the magnetic theory (dual SQCD)}

Finally we consider the Banks-Zaks expansion for the critical
exponent in the dual SQCD theory,
\beq
\gamma=G_1^d\delta_d^2+G_2^d\delta_d^3+G^d_3\delta_d^4+\cdots
\label{gamma_dual_SQCD_BZ}
\eeq
There are two ways to calculate this quantity, one, which has been
used in \cite{Anselmi}, is based directly 
on the coupled renormalization group equations (\ref{coupled_RG}) and
the other is based on the reduced equation (\ref{reduced_RG}).
We show that both methods give the same Banks-Zaks expansion. 

Calculating $\gamma$ in the magnetic theory directly from the coupled
renormalization-group equations (\ref{coupled_RG}) is more involved,
since there are two couplings.  As mentioned above, a similar calculation was 
performed in \cite{Anselmi}. The latter ref. presents a calculation of
the leading-order term in the expansion, but in fact, as we shall see,
the 2-loop gauge  $\beta$
function together with the one-loop Yukawa $\beta$ function
fixes also the next-to-leading order term, just like in QCD
and in the SQCD electric theory.  

Let us briefly describe the method and then give the results. 
The generalization of $\gamma$ to a two coupling theory is the 
following matrix:
\begin{eqnarray}
\label{gamma_mat_def}
\Gamma\,=\,\left.\left(
\begin{array}{cc}
{\displaystyle \frac{d\beta^d_x}{dx}} & 
{\displaystyle \frac{d\beta^d_x}{d\lambda}} \\
\\
{\displaystyle \frac{d\beta^d_{\lambda}}{dx}} &
{\displaystyle \frac{d\beta^d_{\lambda}}{d\lambda}} \\
\end{array}
\right)\right|_{\FP}
\end{eqnarray}
The next step is to diagonalize the matrix. This yields two
eigenvalues: $\gamma_1$ and $\gamma_2$. Therefore a physical
quantity behaves in the infrared like
\beq
x_{\FP}-x=K_1\ \left(Q^2/\Lambda^2_{\eff}\right)^{\gamma_1}
+K_2\ \left(Q^2/\Lambda^2_{\eff}\right)^{\gamma_2}
\eeq
and then asymptotically only the minimal eigenvalue is important.
Thus we conclude that $\gamma={\rm min}\left\{\gamma_1,\gamma_2\right\}$.

Taking the derivatives of the coupled $\beta$ functions
(\ref{coupled_RG}) at the fixed-point 
we find the matrix elements of (\ref{gamma_mat_def}):
\begin{eqnarray*}
\left.\frac{d\beta^d_x}{dx}\right\vert_{\FP}
&=& {\displaystyle \frac {392}{3}} \,{\displaystyle 
\frac {N_c^{2}}{N_c^{2} - 4}} \,\delta_d^2 \,+\, 
{\displaystyle \frac {1120}{9}} \,{\displaystyle \frac {N_c
^{2}\,(4 + 13\,N_c^{2})}{(N_c^{2} - 4
)^{2}}}\,\delta_d^3 \,+\,\cdots
\\
\\
\left.\frac{d\beta^d_x}{d\lambda}\right\vert_{\FP}&=& - 784\,{\displaystyle 
\frac{N_c^{4}}{(N_c^{2} - 4)^{2}}} \,\delta_d^2
\,-\, {\displaystyle \frac {448}{3
}} \,{\displaystyle \frac {N_c^{4}\,( - 36 + 65\,N_c^{2
})}{(N_c^{2} - 4)^{3}}} \,\delta_d^3\,+\,\cdots
\\
\\
\left.\frac{d\beta^d_{\lambda}}{dx}\right\vert_{\FP}&=&
-{\displaystyle \frac {4}{3}} \,{\displaystyle 
\frac {(N_c^{2} - 4) }{N_c^{2}}}\,  \delta_d
\,-\, {\displaystyle 
\frac {8}{9}} \,{\displaystyle \frac {(13\,N_c^{2} + 28)
}{N_c^{2}}} \,\delta_d^2\,+\,\cdots
\\
\\
\left.\frac{d\beta^d_{\lambda}}{d\lambda}\right\vert_{\FP}&=& 
{\displaystyle \frac {28}{3}} \, \delta_d \,+\, 
{\displaystyle \frac {8}{9}} 
\,{\displaystyle \frac {(79\,N_c^{2} + 76)\,}{N_c^{2} -
4}}\,\delta_d^2 \,+\,\cdots
\end{eqnarray*}
The eigenvalues are 
\begin{eqnarray}
\label{two_eigenvalues}
\gamma_1&=& \frac{28}{3}\,\delta_d
\,+\,\frac{8}{9}\frac{205N_c^2+76}{N_c^2-4}\,\delta_d^2
\,+\,{\cal O}(\delta_d^3)\\ \nonumber
\gamma_2&=& \frac{56}{3}\frac{N_c^2}{N_c^2-4}\,\delta_d^2\,-\,
\frac{64}{9}\frac{(17N_c^2+2)N_c^2}{(N_c^2-4)^2}\,\delta_d^3\,+\,
{\cal O}(\delta_d^4)
\end{eqnarray}
The two eigenvalues are positive reflecting the infrared stability of 
the fixed point. The smaller eigenvalue is $\gamma=\gamma_2$.   

We note that these eigenvalues do not agree with the leading order
calculation in \cite{Anselmi}. 
The reason\footnote{The authors are in debt to D. Anselmi for his help on
this matter.} is that ref. \cite{Anselmi} uses the $\beta$ function as
it appears in \cite{KSV} -- see the comment concerning \cite{KSV} 
below eq. (\ref{dual_coef}).    

The second method to calculate $\gamma$ in the magnetic theory is to
use the reduced renormalization group equation. Here we have
a single coupling and thus the derivative of the $\beta$ function
(\ref{reduced_RG}) at the fixed-point immediately yields the relevant
$\gamma$. Performing this calculation we indeed find the same value as
appears in (\ref{two_eigenvalues}) for $\gamma_2$. 
The non-relevant perturbation
corresponding to $\gamma_1$ does not even appear when using this procedure.

As stressed above the critical exponent can be (and was) calculated up
to the next to leading order term from the available coefficients in the
magnetic theory $\beta$ function. This conclusion is transparent in the
reduction method: the 2-loop reduced $\beta$ function is fully determined
from the 2-loop gauge $\beta$ function and the 1-loop Yukawa $\beta$
function. In particular it does not depend on the 2-loop Yukawa $\beta$
function. Like in QCD and in the electric SQCD cases, it follows that
$\gamma$ can be computed up to the next to leading order term.
On the other hand, this conclusion is non-trivial when calculating
$\gamma$ from the coupled $\beta$ functions. It turns out that the next to
leading order terms in the {\em matrix elements} of $\Gamma$, likewise
in the {\em larger} eigenvalue $\gamma_1$, do depend on the (unknown) 
2-loop terms of the Yukawa coupling $\beta$ function. On the other hand,
the relevant eigenvalue $\gamma_2$ does not depend on these 2-loop
terms.

We emphasize that the first method described above, i.e. to use the coupled
$\beta$ function in order to define $\gamma$ as a matrix, and then take the
minimal eigenvalue, is a completely general procedure.  
It is guaranteed that the second method that uses the reduced
$\beta$ function will also give the correct value of $\gamma$
once we choose
the reduced solution that belongs to the infrared {\em stable}
fixed-point, which we did. 
Note that there exists in this case another possible reduction
that corresponds to $\lambda=0$ \cite{Oehme_reduction,ee_duality} and an
{\em unstable} infrared fixed-point. Since
$\lambda=0$ means a zero superpotential, this reduction cannot
correspond to the dual of an electric theory. 
Had we used this reduction instead of the relevant one we would have
obtained different values for the infrared fixed-point and for $\gamma$.  

Contrary to QCD and the electric SQCD
theory, higher order corrections to the coupled $\beta$ functions of
the magnetic SQCD theory are not known. 
As a result we cannot study the behavior and the break-down of the 
Banks-Zaks series beyond the next-to-leading order term in this case.

\subsection{Interpolating between the original theory and its dual}

The results of the previous sections indicate that in SQCD the
electric theory is strongly coupled when the magnetic theory is weakly 
coupled and vise-versa. These results are in accordance with
Seiberg's description of the conformal window.
Since calculations are usually limited to the weak coupling regime,
it is in general impossible to compare between results obtained in the two
theories. An interesting example of how the two theories can be compared for
the infrared limit of a specific physical quantity, 
the total ``hadronic'' cross section ratio in $e^+e^-$ 
annihilation, $R_{e^+e^-}$, is discussed in ref. \cite{ee_duality}.

We find it useful to express
the relation between the infrared limit of physical quantities in
terms of the relation between the corresponding effective charges. 
For instance,
we can define an effective charge in the original (electric) theory,
\beq 
R_{e^+e^-}(Q^2)\equiv a+b\,x_{R_{e^+e^-}}(Q^2),
\eeq 
where $a$ and $b$ are $N_c$ and $N_f$ dependent constants, 
and another effective charge in the dual (magnetic) theory, 
\beq
R^d_{e^+e^-}(Q^2)\equiv a_d+b_d \,x^d_{R_{e^+e^-}}(Q^2).
\eeq
For a non-zero $Q^2$ the quantities in the two theories 
are not related, but in the
infrared limit duality relates them. Close to the fixed-point we have:
\begin{eqnarray}
x_{R_{e^+e^-}}(Q^2)&=&x_{R_{e^+e^-}}(0)-
\left(\frac{Q^2}{\Lambda_{\eff}^2}\right)^\gamma \\ \nonumber
x^d_{R_{e^+e^-}}(Q^2)&=&x^d_{R_{e^+e^-}}(0)-
\left(\frac{Q^2}{\Lambda^2_{\eff_d}}\right)^{\gamma_d} 
\end{eqnarray}
According to duality, we have in the infrared limit,
$R^d_{e^+e^-}(0)=R_{e^+e^-}(0)$, i.e.
\beq
a+b\,x_{R_{e^+e^-}}(0)=a_d+b_d\,x^d_{R_{e^+e^-}}(0)
\label{Ree}
\eeq
and 
\beq
\gamma=\gamma_d.
\eeq
The last equality was shown to follow from duality in \cite{Anselmi} 
where $\gamma$ was identified as the anomalous dimension of the 
Konishi current, which is part of the superconformal algebra.
A natural extension of duality would be to conjecture
\cite{preparation} that also the
terms that describe the {\em approach} to the infrared fixed-point are
the same in the two dual theories. In this case it makes sense to
set the convention such that $\Lambda_{\eff}=\Lambda_{\eff_d}$.

From Seiberg's description of the conformal window it seems reasonable
to assume that taking into account the physical information
from both the electric and magnetic descriptions together, 
we may be able to describe the infrared limit in the entire window. 
In the following we give an example how this can be achieved in
practice by
interpolating between the Banks-Zaks expansions in the electric and
magnetic theories.

The quantity we consider is the critical exponent $\gamma$
which was calculated in the original and the dual theories in the
previous section. As explained above, the calculations in the two
theories correspond to {\em the same physical quantity}.

We choose to analyze the critical exponent in the large $N_c$
limit (a similar analysis is possible for any $N_c\geq3$). 
Using the expansions for $\gamma$ in electric theory (\ref{gamma_SQCD_BZ})
and in the magnetic theory $\gamma=\gamma_2$ (\ref{two_eigenvalues}),
we obtain in the large $N_c$ limit the following partial-sums, respectively:
\begin{eqnarray}
\label{gamma_infty}
\gamma&=&\frac{1}{6} \left[3-R\right]^2+\frac{1}{9} \left[3-R\right]^3
+\left(-\frac{1}{12} \zeta_3+\frac{1}{216}\right) \left[3-R\right]^4+\cdots \\ 
\gamma&=&
\frac{56}{3} \left[R-\frac{3}{2}\right]^2-
\frac{1088}{9} \left[R-\frac{3}{2}\right]^3+\cdots \nonumber
\end{eqnarray}
Fig.~\ref{SQCD_gamma_interpolation} presents 
the functional form of $\gamma$, according to 
the above expansions.
In the original theory we show 
\hbox{$G_1\delta^2+G_2\delta^3$} and 
\hbox{$G_1\delta^2+G_2\delta^3+G_3\delta^4$},
and in the dual theory we show \hbox{$G_1^d\delta_d^2+G_2^d\delta_d^3$}.
Each of the two expansions can be trusted just in some
limited region around the expansion point. It is quite clear from  
this figure, especially if one compares the results obtained 
in both descriptions,
that the perturbative result cannot be extrapolated to
the far-end of the conformal window.
Thus, a straightforward comparison between the results obtained in the
two dual description is impossible. On the other hand, it makes sense
to interpolate between the two. 

We use the 2-point Pad\'e approximants method to interpolate between
the two series (\ref{gamma_infty}). The general idea is to
construct a rational function which yields both the known series
(\ref{gamma_infty}) in the original and dual theories
when expanded in a Taylor series at $R=3$ and $R=3/2$, respectively.
The calculation technique is explained in detail in chapter 
8 in \cite{Baker}. The resulting approximant is: 
\begin{eqnarray}
\label{gamma_PA}
\gamma_{\PA}=
\displaystyle{
\frac{392\, [3-R]^2\,[R-(3/2)]^2}
{\begin{array}{l}
\left[
23814\,\zeta_3 + 10881 - (47628\,\zeta_3 - 22734)\,
R + (34398\,\zeta_3 + 17685)\,R^{2}\right. \\
\left.
\,\,\,\,\,\,\,\,\,\,\,\,\,\,\,\,\,\,\,\,\,\,\,\,\,\,\,\,\,\,\,\,\,
\,\,\,\,\,\,\,\,\,\,\,\,\,\,\,\,\,\,\,\,\,\,\,\,\,\,\,
- (10584\,\zeta_3 + 6216)\,R^{3} 
+ (1176\,\zeta_3 + 880)\,R^{4}\right]
\end{array}}}
\end{eqnarray}
This approximant is a rational polynomial of order [4/4]. 
Note that the numerator in (\ref{gamma_PA}) contains the double zero at both
ends of the conformal window as implied by (\ref{gamma_infty}).
In principle, there could be further possible [N/M] approximants, based on
other rational polynomials. But, given the particular form of
(\ref{gamma_infty}), other approximants cannot be constructed at this
order.
        
The interpolating $\gamma_{\PA}$ of (\ref{gamma_PA})
is shown in fig.~\ref{SQCD_gamma_interpolation} 
together with the lines describing the partial-sums (\ref{gamma_infty})
which correspond to the Banks-Zaks expansions around $R=3$ and $R=3/2$. 
We stress that in order to construct (\ref{gamma_PA}) we used nothing
but the information contained in the coefficients of
(\ref{gamma_infty}). A priori, a Pad\'e pole could have appeared within the
conformal window, which would probably mean that this interpolation
technique is inappropriate. We find that such a pole does not
appear. On the other hand, the convergence radius of the $\delta$ and
$\delta_d$ expansions of (\ref{gamma_PA}) are rather small due to 
complex Pad\'e poles. This should not be a surprise, as we expect the
all order result to have zero radii of convergence in terms of
$\delta$ and $\delta_d$.   

Another interesting observation is that duality suggests an alternative way
to define the expansion parameters for both the original and the dual
theory. Referring to the original degrees of freedom, we defined
$R\equiv N_f/N_c$, and then the expansion parameters were: in the original
theory $\delta\equiv 3-R$, which is small at the top of the window
and in the dual theory $\delta_d\equiv R-(3/2)$, which is small at the
bottom of the window. If we instead start with the dual theory, then we
consider the ratio
\beq
\tilde{R}\equiv N_f/N_c^d=\frac{N_f}{N_f-N_c}=\frac{R}{R-1} 
\eeq
and define the following expansion parameters:
\begin{eqnarray}
\label{delta_tilde}
\tilde{\delta}&\equiv&\tilde{R}-\frac{3}{2}=\frac{\delta}{4-2\delta}\\
\nonumber
\tilde{\delta_d}&\equiv&3-\tilde{R}=\frac{4\delta_d}{1+2\delta_d}.
\end{eqnarray}
It is then possible to repeat the calculation of the Banks-Zaks
coefficients in terms of $\tilde{\delta}$ in the electric theory and
$\tilde{\delta_d}$ in the magnetic theory. An alternative way to
calculate the expansion in terms of $\tilde{\delta}$ and
$\tilde{\delta_d}$ would be simply to use the expansion of
(\ref{gamma_infty}):
substituting $\delta$ and $\delta_d$ in terms of $\tilde{\delta}$ and
$\tilde{\delta_d}$, according to (\ref{delta_tilde}) 
in (\ref{gamma_infty}) and Taylor
expanding to the maximal order to which the coefficients are fixed 
(order $\tilde{\delta}^4$ in the electric theory and order
$\tilde{\delta}_d^3$ in the magnetic theory) we find:
\begin{eqnarray}
\label{gamma_infty_tilde}
\gamma&=&\frac {8}{3} \, \left[\tilde{R}-\frac32\right]^2
- \frac {32}{9}\left[\tilde{R}-\frac32\right]^3
+ \left( - \frac {256}{27}  - \frac {64}{3} \,\zeta_3\right)
\left[\tilde{R}-\frac32\right]^4+\cdots \\ 
\gamma&=&
\frac {7}{6} \left[3-\tilde{R}\right]^2
-\frac {13}{18} \left[3-\tilde{R}\right]^3+\cdots \nonumber
\end{eqnarray}

Note that the functional form and therefore the numerical values of the
truncated expansions for
$\gamma$ using the $\tilde{\delta}$ and $\tilde{\delta}_d$ variables 
is {\em different} (at any finite order) from those using the 
$\delta$ and $\delta_d$ variables. On the other hand, $\gamma$ is a
physical quantity and therefore it cannot depend on such an arbitrary
choice of expansion parameter. This is a disadvantage of using 
truncated series: they do not respect the invariance property of the
full function $\gamma(N_f/N_c)$ to the choice of expansion parameter.

Finally, we come back to the 2-point Pad\'e Approximant
(\ref{gamma_PA}) and ask how 
this function behaves under this change of expansion parameter. One can
construct a [4/4] 2-point Pad\'e approximant starting from the series
(\ref{gamma_infty_tilde}) and compare it, as a function of $N_f/N_c$ 
to the [4/4] 2-point Pad\'e approximant of (\ref{gamma_PA}). One would
find that the two functions are {\em identical}. The reason is that
the transformation relating $R$ to $\tilde{R}$, and then also the
transformations relating $\delta$ and $\delta_d$ to $\tilde{\delta}$ and
$\tilde{\delta}_d$, are all homographic transformations of the argument of
the Pad\'e approximant. It is then guaranteed by a mathematical
theorem (see \cite{Baker,Why}) for each expansion separately, 
that the diagonal Pad\'e function is invariant. Note that this theorem
holds in our example which is a diagonal [4/4] rational function, but
it does not hold for non-diagonal rational polynomials. 
It is interesting to mention that the same invariance property of
Pad\'e approximants was shown to be significant in a 
different context \cite{Why}, where Pad\'e approximants are used to
resum perturbative series in QCD. 

This invariance property of the diagonal 2-point Pad\'e approximants 
with respect to the choice of expansion parameter, indicates that it 
is a good candidate to serve as an approximation to the function 
$\gamma(N_f/N_c)$.    

In order to have a rough estimate of the accuracy of this
approximation procedure,
we suggest the following: let us construct a higher approximant based
on a guess for the higher order coefficients of the expansion. We
choose to ``add'' a coefficient in the Banks-Zaks expansion in the
magnetic theory, $G_3^d$. Not knowing anything about higher order
coefficients of the coupled $\beta$ functions that are needed
to determine this coefficient, we just use $G_3^d=0$. We do not expect
it to be a good guess (especially considering the presumably 
asymptotic nature of the expansion) but we use it just to check 
the sensitivity of the calculation. 
With a value of $G_3^d$ at hand, we can obtain a (non-diagonal!) [4/5]
rational polynomial that reproduces the first three terms in each expansion:
the first three correct coefficients in the electric theory, and the first
two correct coefficients in the magnetic theory, with the third
coefficient put to zero. The resulting [4/5] approximant is also shown
in fig.~\ref{SQCD_gamma_interpolation}. We refer to the difference
between the [4/4] approximant of (\ref{gamma_PA}) and the latter as a
rough estimate of the error of the interpolation which is due to the lack of
knowledge of higher order corrections.  

This exercise suggests that although the partial-sums 
can be considered to be a good numerical
approximation to the physical quantity $\gamma$ only in some limited
domain around the boundaries of the conformal window, an interpolating Pad\'e
approximant can be a good approximation for $\gamma$ in the entire 
conformal window. 

Unfortunately, in the absence of non-perturbative calculations, it is
impossible to validate duality, nor to check our prediction for
$\gamma$ in the interior of the conformal window.
This calculation can
surely be improved if higher-order terms in the $\beta$
function of either the electric or the magnetic theory will be
available. It is of course possible that there exist
non-perturbative corrections that limit in principle 
the accuracy of this calculation. One should note, however, that
although the Banks-Zaks expansions we started with (\ref{gamma_infty})
are presumably asymptotic series, with zero radius of convergence, the
resulting function (\ref{gamma_PA}) is well-defined in the entire
conformal window. Thus combining the information from the two theories
in this 2-point Pad\'e method, is in fact also a way to resum the
divergent series. 

If there were many known terms in the electric Banks-Zaks
expansion one could use some resummation technique of this single
series (such as Borel resummation or the application of an 
ordinary Pad\'e approximant),
hoping to get a vanishing value for $\gamma$ at the other side of the
conformal window, thus adding a further consistency check to the
duality conjecture. 

It is important to stress that if there was a phase transition
somewhere within the conformal window, controlled by non-perturbative
effects that are inaccessible from neither side of the conformal
window, our interpolation
procedure would not have been meaningful. However, the phase structure
picture drawn in \cite{Seiberg} suggests a
completely {\em smooth} transition between the top of the conformal window
where the electric theory is weakly coupled and the bottom of the
window where the dual theory is weakly coupled.

We emphasize that the interpolation for $\gamma$ serves here
just as an example, and the method can be applied in general to any 
{\em physical quantity} that can be calculated in perturbation theory
in {\em both} the original and the dual theories, such as the one in
eq. (\ref{Ree}).

\section{Conclusions}

The purpose of this paper is to understand the nature of the
non-trivial infrared fixed-point that appears in asymptotically free
theories such as QCD and SQCD if the number of light flavors is large enough. 
The main question we deal with is whether this fixed-point
always originates within perturbation theory, or is it due to
non-perturbative physics. 

Since more is known on the phase structure of SQCD, it is natural to
discuss this theory first. 
The argument for the presence of a fixed-point in SQCD in
\cite{Seiberg}, just like in QCD \cite{BZ}, is a purely 
perturbative one based on the 2-loop $\beta$ function.  
On the other hand, Seiberg's picture of the 
conformal window \cite{Seiberg} {\em assumes} that the electric 
theory is strongly coupled near the lower boundary of the conformal 
window, and thus the fixed-point there is non-perturbative. 
Therefore, our first task dealing with the SQCD case was to confirm that the
electric theory is strongly coupled at large distances near the bottom
of the conformal window. Indeed we showed by considering the
analyticity structure of the coupling and by examining the Banks-Zaks
expansion, that the presence of the fixed-point at the far-end of the 
conformal window cannot be established in perturbation theory: it is
a non-perturbative fixed-point. 
Our investigation further shows that the perturbative analysis signals
its own inapplicability to describe the infrared already before the
dual theory becomes weakly coupled. 

In order to describe the infrared physics in the entire conformal
window it is useful to combine the
information from perturbative calculations in both the electric and
magnetic descriptions. We suggest to
use an interpolation method -- a 2-point Pad\'e approximant -- 
that combines the Banks-Zaks expansions  around the two ends of the
conformal window 
into a single formula. This method was demonstrated in Sec. 4.3 for the
critical exponent, but in fact it is applicable to any physical quantity
which is calculable in perturbation theory in both theories.  
The invariance property of the diagonal 2-point Pad\'e approximant under
the change of expansion parameter (a change which is motivated by
the symmetrical realization of duality in the conformal window) 
makes it appropriate to describe $\gamma(N_f/N_c)$.
Still, this is of course not an exact calculation: it is quite clear that
further corrections, both such that are accessible by
perturbation theory in either the electric description or the magnetic
one, and eventually such that are not, do exist.

\begin{figure}[htb]
\begin{center}
\mbox{\kern-0.5cm
\epsfig{file=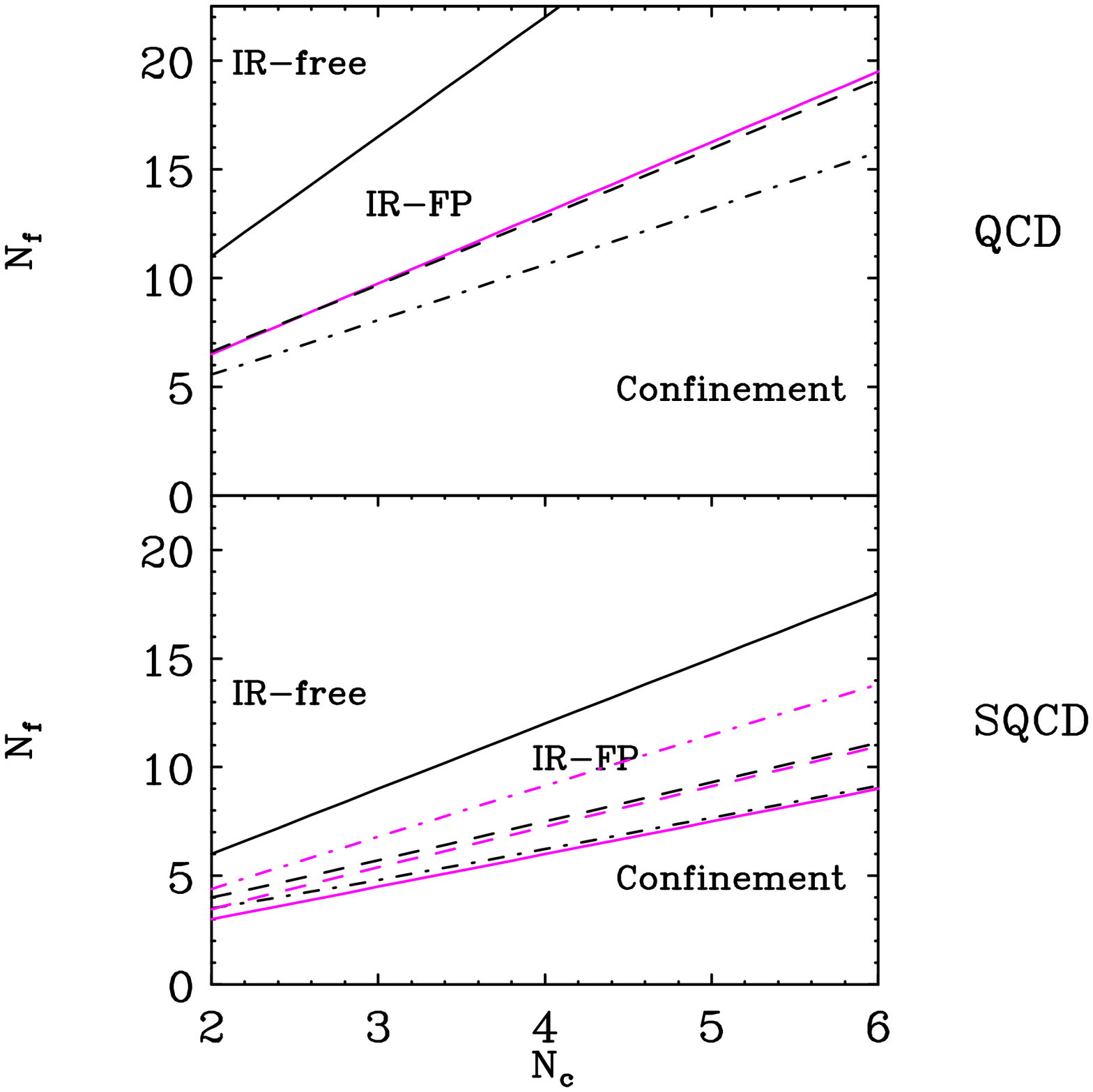,width=15.0truecm,angle=90}
}
\end{center}
\caption{The conformal window in QCD (upper plot) and SQCD (lower
  plot) is shown in the $N_f\,\, -\,\, N_c$ plane. In both plots the
  $\beta_0=0$ line, separating the infrared free phase from the
  ultraviolet asymptotically free phase is drawn as a continuous black
  line. This line is the upper boundary of the conformal window. 
  The lower boundary of the conformal window as implied by
  superconvergence is drawn in gray. In the SQCD case, this last
  line is also the line below which the dual theory becomes infrared
  free. In both plots, 
the (black) dashed line shows the lower boundary of the region
  where the 2-loop coupling has a causal analyticity structure. Below
  this line and above the dot-dash line there are complex Landau
  singularities. Below the dot-dash line there is a space-like Landau
  branch point. In the lower plot, we also show in gray the dual lines which
  describe the analyticity structure of the dual coupling constant.}
\label{conformal_window}
\end{figure}

\begin{figure}[htb]
\begin{center}
\mbox{\kern-0.5cm
\epsfig{file=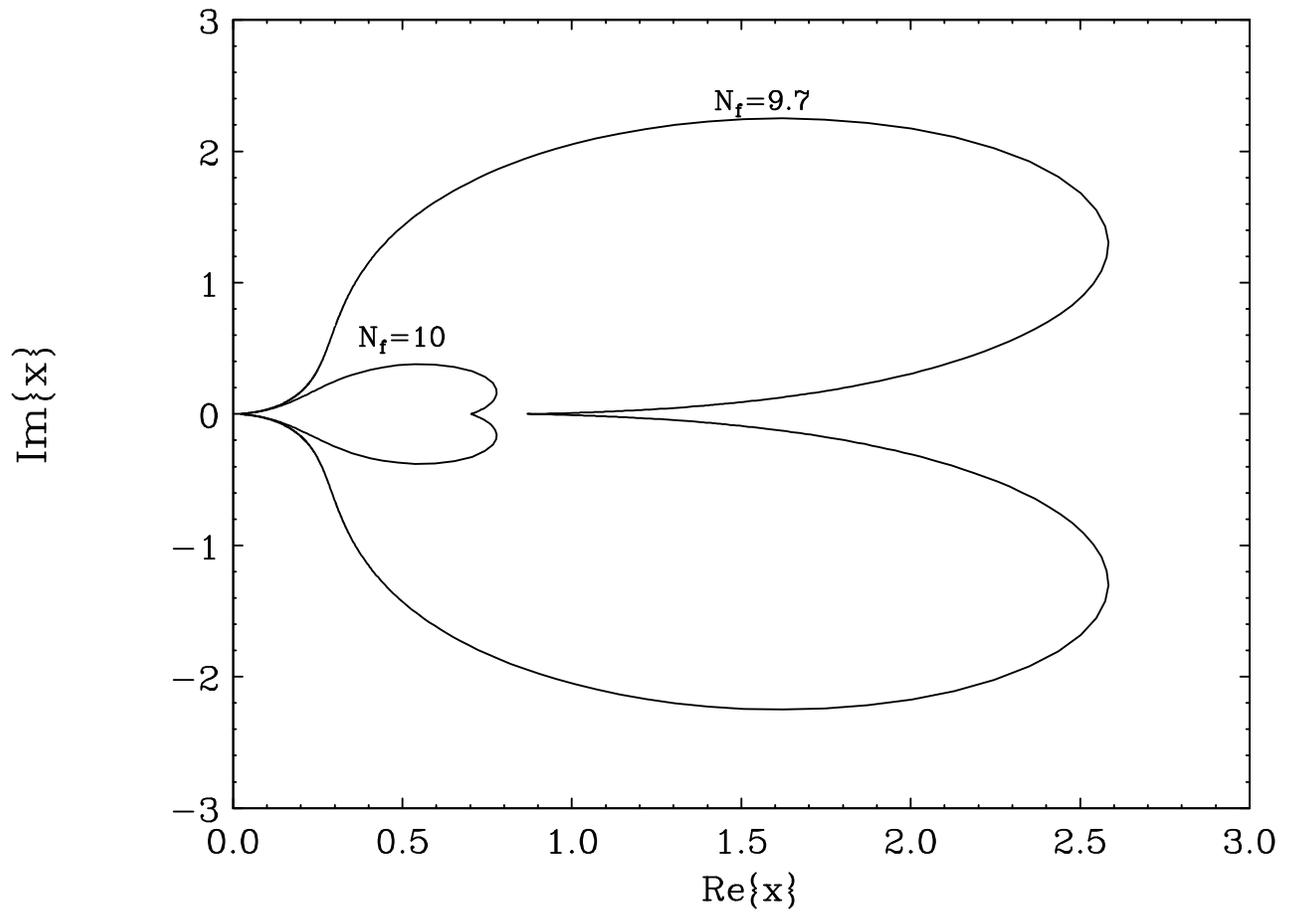,width=12.0truecm,angle=90}
}
\end{center}
\caption{The compact domain in the complex coupling plane which
corresponds to the entire complex $Q^2$ plane according to the 2-loop
$\beta$ function in $N_c=3$ QCD with $N_f=10$ and $N_f=9.7$, i.e. just
above the minimal value of $N_f$ required for causality of the 2-loop coupling 
($N_f\simeq 9.683$).
 }
\label{x_plane_Nc3}
\end{figure}

\begin{figure}[htb]
\begin{center}
\mbox{\kern-0.5cm
\epsfig{file=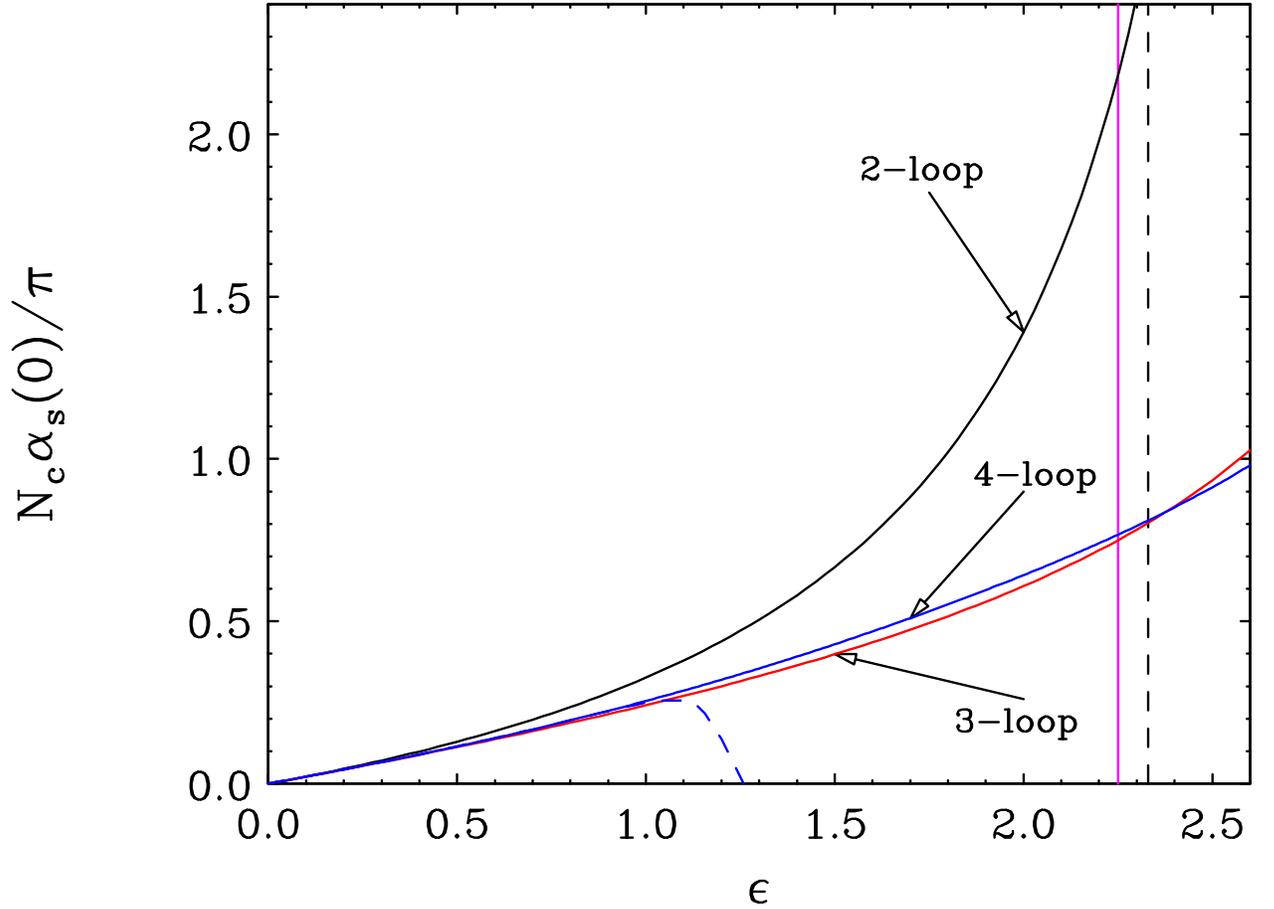,width=12.0truecm,angle=90}
}
\end{center}
\caption{The infrared limit of the 2-loop coupling and the 3-loop and
4-loop ${\hbox{$\overline{\hbox{MS}}\,$}}$ couplings in large $N_c$
QCD as a function of $\epsilon\equiv (11/2)-(N_f/N_c)$.
The dashed line corresponds to the 20th order 
Taylor expansion in $\epsilon$ of the 4-loop coupling. 
The continuous vertical line represents the bottom of the conformal window
implied by superconvergence and the dashed vertical line shows the
2-loop causality boundary.
 }
\label{QCD_IR_epsilon}
\end{figure}

\begin{figure}[htb]
\begin{center}
\mbox{\kern-0.5cm
\epsfig{file=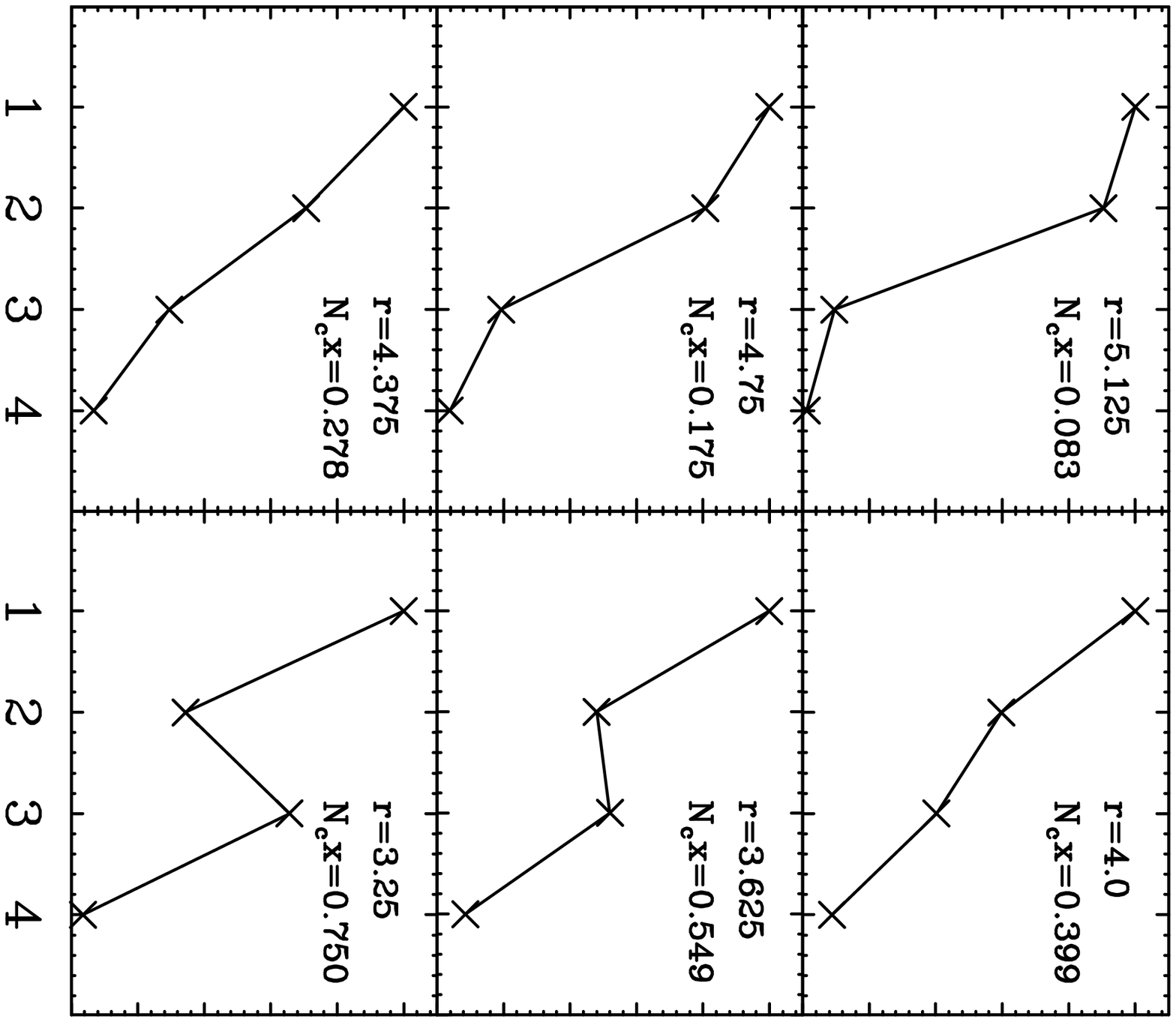,width=12.0truecm,angle=90}
}
\end{center}
\caption{The relative magnitude of the four leading terms in the 
 large $N_c$ QCD $\beta$
 function 
 in the ${\hbox{$\overline{\hbox{MS}}\,$}}$ renormalization
 scheme in the infrared limit, for various values of
 $r=N_f/N_c$, from the top of the conformal window ($r=5.5$) down to 
the bottom ($r=3.25$).  
The normalization in each plot is such that the leading order term is
 1. This means that the second is: $\beta_1x/\beta_0$, the third:
 $\beta_2x^2/\beta_0$ and the fourth: $\beta_3x^3/\beta_0$.  
The value of the coupling $x$ is
 calculated as an explicit solution of the equation $\beta(x)=0$,
 with the 3-loop $\beta$ function. Using the coupling as a zero of the
 4-loop $\beta$ function does not change much the results.
 }
\label{QCD_terms_}
\end{figure}

\begin{figure}[htb]
\begin{center}
\mbox{\kern-0.5cm
\epsfig{file=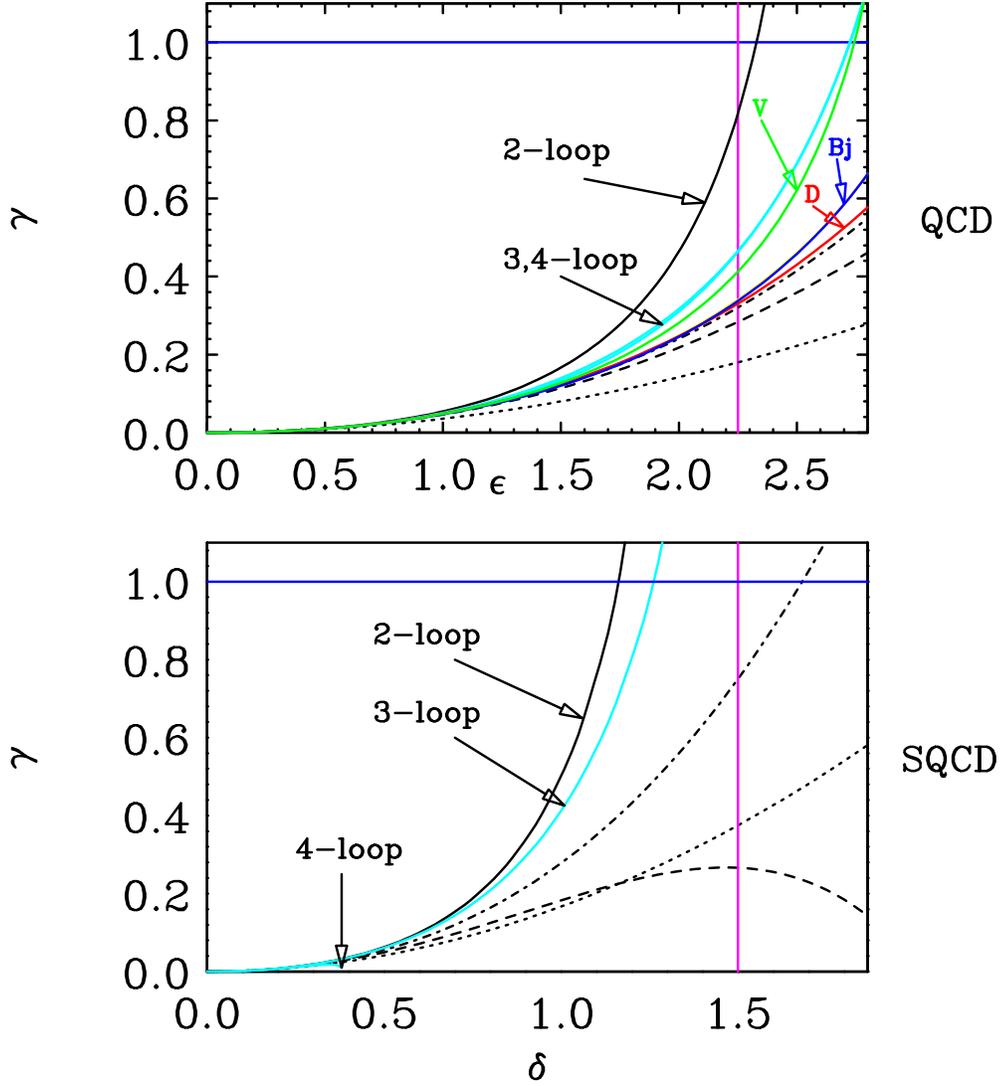,width=14.5truecm,angle=90}
}
\end{center}
\caption{The critical exponent $\gamma$ in large $N_c$ QCD (SQCD) is
shown in the upper (lower) plot as a function of
$\epsilon=(11/2)-(N_f/N_c)$ ($\delta=3-(N_f/N_c)$), 
according to an explicit calculation 
from the truncated 2-loop $\beta$ function and the 3-loop and 4-loop
$\beta$ functions (the loop order is indicated by the arrows) 
in the ${\hbox{$\overline{\hbox{MS}}\,$}}$ (DRED) 
renormalization scheme as well as according to the three available 
partial sums in the Banks-Zaks expansion: leading order -- dotted line,
next-to-leading order -- dot-dashed line, and next-to-next-to-leading
 order -- dashed line. In the QCD plot we show also the explicit
calculation of $\gamma$ at the 3-loop order in various physical
renormalization schemes defined from the vacuum polarization 
D-function (D), the polarized and non-polarized Bjorken sum-rules (Bj)
and the heavy quark effective potential (V). 
The vertical arrow in the SQCD plot shows the point where the 
positive real zero of the 4-loop $\beta$ function disappears.
The vertical line is the bottom of the conformal window
 according to superconvergence and the horizontal 
$\gamma=1$ line is the necessary upper bound of $\gamma$ for a causal
analyticity structure.}
\label{combined_gamma_BZ}
\end{figure}

\begin{figure}[htb]
\begin{center}
\mbox{\kern-0.5cm
\epsfig{file=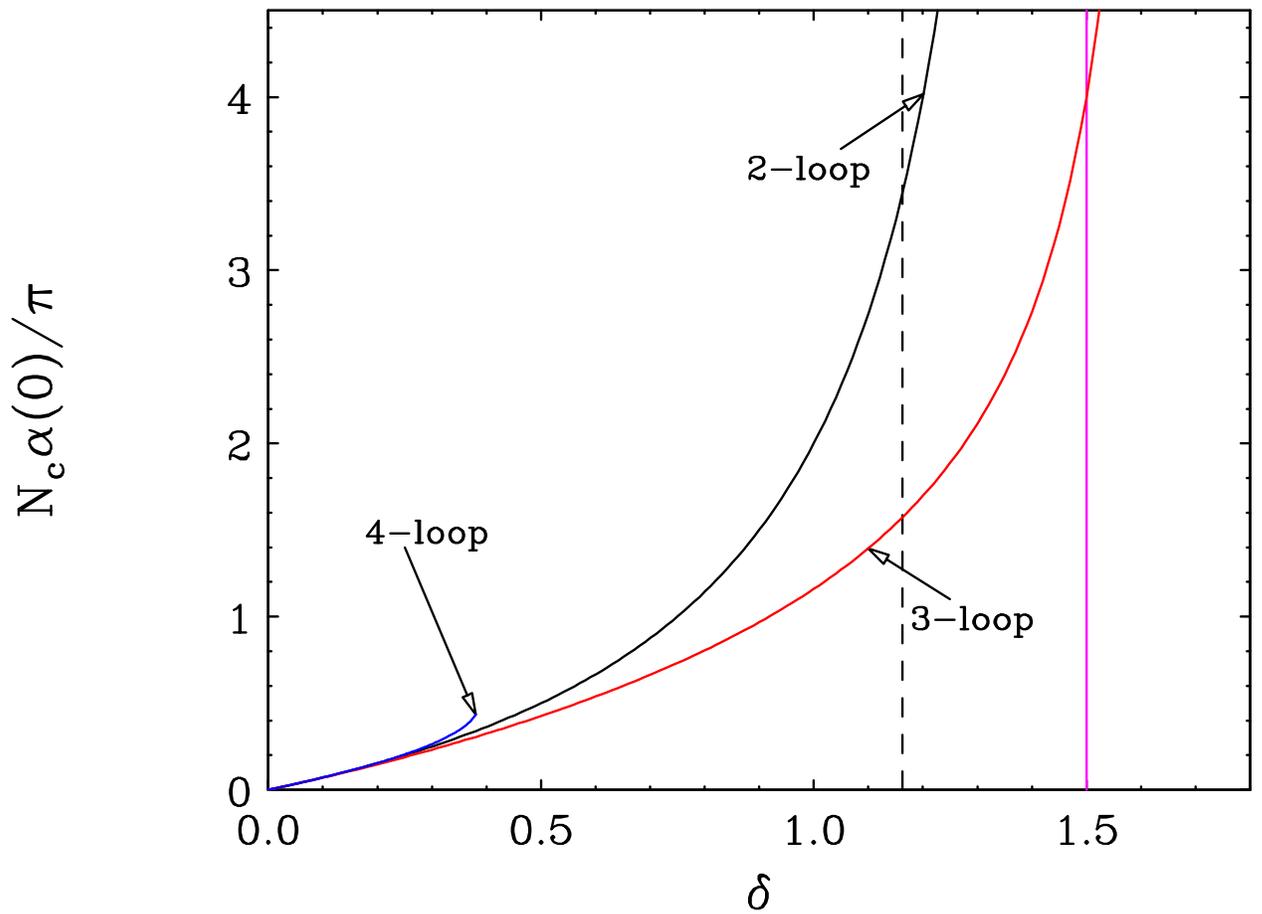,width=12.0truecm,angle=90}
}
\end{center}
\caption{The infrared limit of the 2-loop coupling and the 3-loop and
4-loop DRED couplings in large $N_c$ SQCD 
as a function of $\delta\equiv 3-(N_f/N_c)$.
The continuous vertical line represents the bottom of the conformal window
implied by superconvergence (or duality) and the dashed vertical line shows the
2-loop causality boundary.
 }
\label{SQCD_IR_delta}
\end{figure}

\begin{figure}[htb]
\begin{center}
\mbox{\kern-0.5cm
\epsfig{file=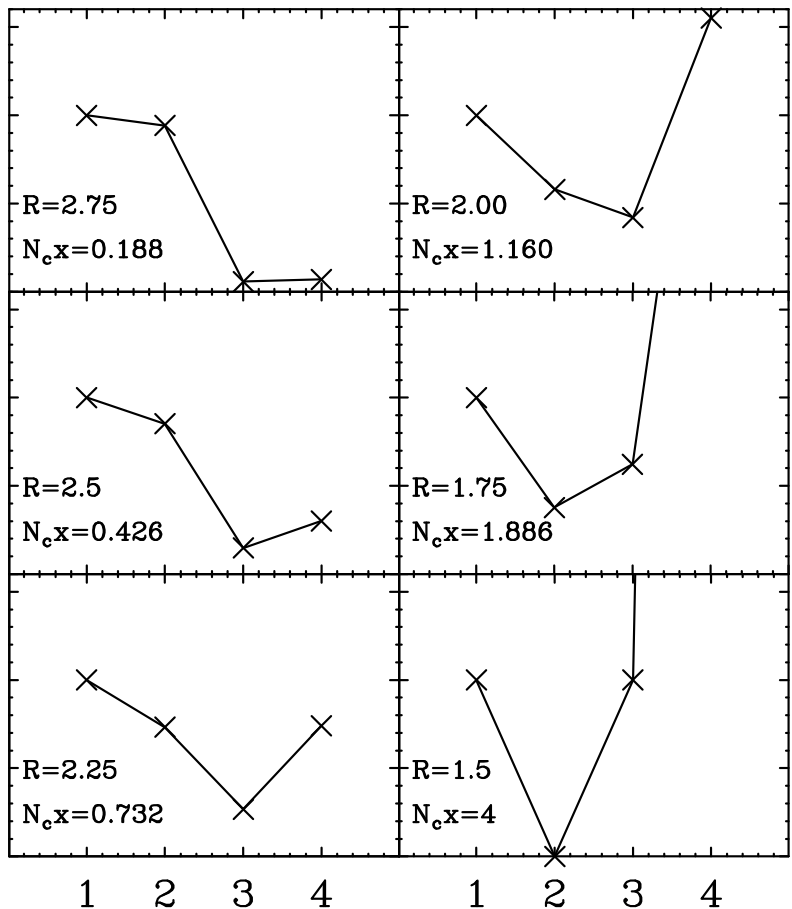,width=12.0truecm,angle=90}
}
\end{center}
\caption{The relative magnitude of the four leading terms in the 
 large $N_c$ SQCD $\beta$
 function 
 in the DRED renormalization
 scheme in the infrared limit for various values of
 $R=N_f/N_c$, from the top of the conformal window ($R=3$) 
 down to the bottom ($R=1.5$).  
The normalization in each plot is such that the leading order term is 1. 
This means that the second is: $B_1x/B_0$, the third:
 $B_2x^2/B_0$ and the fourth: $B_3x^3/B_0$.  
The value of the coupling $x$ is
 calculated as an explicit solution of the equation $\beta(x)=0$,
 with the 3-loop $\beta$ function. 
}
\label{SQCD_terms}
\end{figure}

The lower boundary of the conformal window in the non-supersymmetric
case is not as well established. If the superconvergence criterion
is only a sufficient condition for confinement and not a necessary
one, the conformal window may be narrower 
than shown in fig.~\ref{conformal_window}  (upper plot). On the other
hand, there are evidence from lattice simulations \cite{lattice_Japan} 
indicating that the confining phase transition occurs at lower $N_f$. 
In spite of these contradicting findings, we refer to the lower
boundary of the conformal window as the one implied by superconvergence.

\begin{figure}[htb]
\begin{center}
\mbox{\kern-0.5cm
\epsfig{file=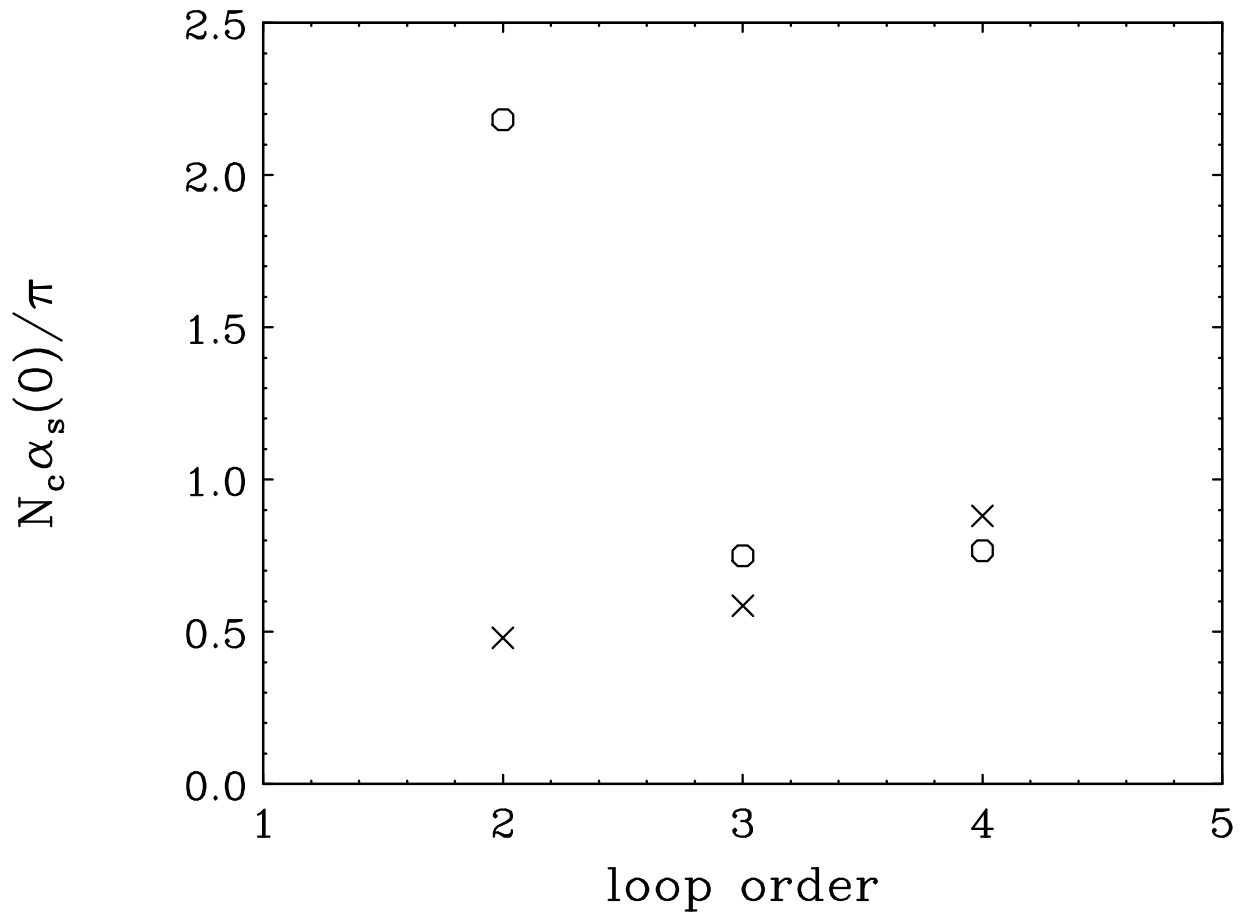,width=12.0truecm,angle=90}
}
\end{center}
\caption{The infrared value of the coupling at the bottom of the conformal
window ($N_f/N_c=13/4$) in large $N_c$ QCD, 
calculated as an explicit solution of the 
equation $\beta(x)=0$ (circles), compared with the corresponding order partial
sum in the Banks-Zaks expansion (crosses). 
The horizontal axis is the loop order of the calculation: 
2, 3 and 4-loop results in ${\hbox{$\overline{\hbox{MS}}\,$}}$ are shown.
 }
\label{QCD_convergence_BZ_loops}
\end{figure}

\begin{figure}[htb]
\begin{center}
\mbox{\kern-0.5cm
\epsfig{file=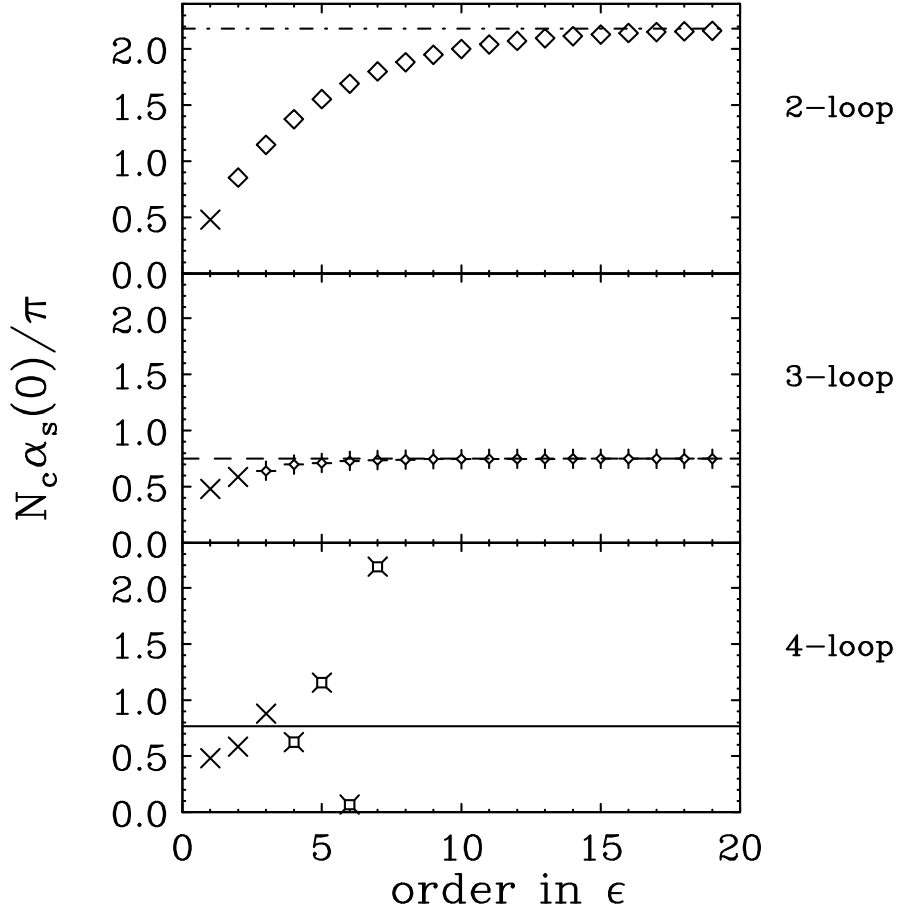,width=12.0truecm,angle=90}
}
\end{center}
\caption{The infrared value of the coupling at the bottom of the conformal
window ($N_f/N_c=13/4$) in large $N_c$ QCD, 
calculated as an explicit solution of the 
equation $\beta(x)=0$. 
Three cases are shown in the three plots, from top to bottom: the
2-loop $\beta$ function and the 3-loop and 4-loop $\beta$ functions in the
${\hbox{$\overline{\hbox{MS}}\,$}}$ renormalization scheme.
The horizontal line is the value of the infared coupling calculated
from $\beta(x)=0$, and the symbols represent the partial sums 
in the expansion in powers of \hbox{$\epsilon\equiv (11/2)-(N_f/N_c)$} of
this solution. The cross symbols 
represent these partial sums that will not be altered by inclusion of
higher order corrections to the $\beta$ function, i.e. they represent
the Banks-Zaks partial sums.
 }
\label{QCD_convergence_BZ_order}
\end{figure}

\begin{figure}[htb]
\begin{center}
\mbox{\kern-0.5cm
\epsfig{file=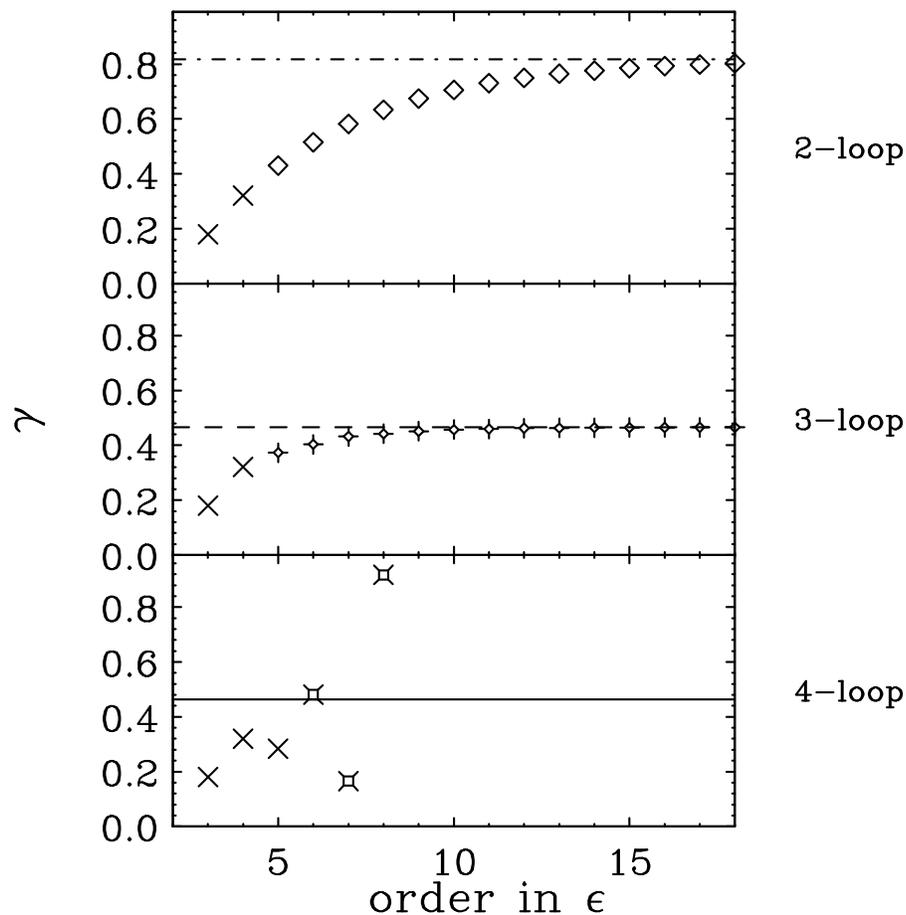,width=12.0truecm,angle=90}
}
\end{center}
\caption{The critical exponent $\gamma$ at the bottom of the conformal
window ($N_f/N_c=13/4$) in large $N_c$ QCD, calculated explicitly from the
truncated $\beta$ function. 
Three cases are shown in the three plots, from top to bottom: the
2-loop $\beta$ function and the 3-loop and 4-loop $\beta$ functions in the
${\hbox{$\overline{\hbox{MS}}\,$}}$ renormalization scheme.
The horizontal line is the value of $\gamma$ calculated
from the truncated $\beta$ function, 
and the symbols represent the partial sums 
in the expansion of $\gamma$ in powers of $\epsilon\equiv (11/2)-(N_f/N_c)$. 
The cross symbols 
represent these partial sums that will not be altered by inclusion of
higher order corrections to the $\beta$ function, i.e. they represent
the Banks-Zaks partial sums which are renormalization
scheme invariant.
 }
\label{QCD_convergence_BZ_gamma_order}
\end{figure}

\begin{figure}[htb]
\begin{center}
\mbox{\kern-0.5cm
\epsfig{file=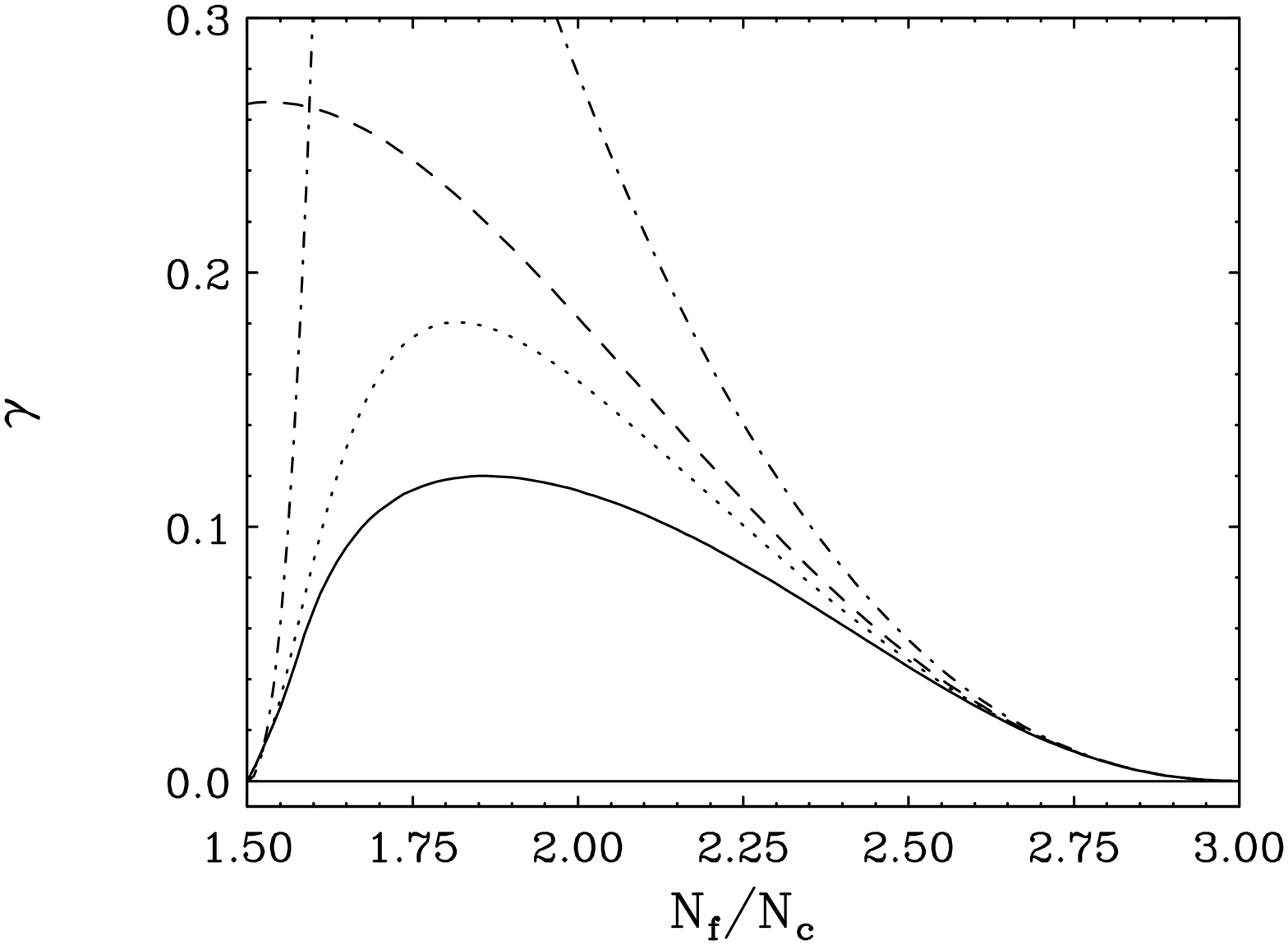,width=12.0truecm,angle=90}
}
\end{center}
\caption{The critical exponent $\gamma$ in large $N_c$ SQCD, 
  is shown as a function of $R=N_f/N_c$ in the
  conformal window.
In the original theory (expansion around $R=3$) we show 
two Banks-Zaks partial sums, according to (\ref{gamma_infty}):
second order partial sum -- dot-dashed line;
third order partial sum -- dashed line.
In the dual theory (expansion around $R=3/2$) we show the second order
  partial sum as a dot-dashed line. The continuous line shows 
the [4/4] interpolating 2-point Pad\'e Approximant of
  eq. (\ref{gamma_PA}). The dotted line shows the [4/5]
  approximant which is based on $G_3^d=0$ and serves as a rough
  measure of the error for the [4/4] interpolating function. 
 }
\label{SQCD_gamma_interpolation}
\end{figure}

Considering the analyticity structure of the coupling and the
Banks-Zaks expansion, we find that perturbation theory in QCD 
is self consistent even at the lower boundary of the conformal
window. The most important evidence supporting this conclusion are: 
\begin{description}
\item{(a) } the 2-loop and 3-loop couplings have a causal analyticity
structure in the entire conformal window. At 3-loop order this is
guaranteed provided $\beta_2<0$, which holds in 
${\hbox{$\overline{\hbox{MS}}\,$}}$ and in all physical schemes 
for which 3-loop coefficients are available. 
\item{(b) } thanks to the small 4-loop term in the
${\hbox{$\overline{\hbox{MS}}\,$}}$ scheme, the standard 
perturbative justification holds down to the infrared limit, and the
3-loop solution for $x_{\MSbar}(Q^2)$ can be trusted in the entire $Q^2$ plane.
\item{(c) } calculation of infrared quantities from the Banks-Zaks
expansion are in reasonable agreement with the explicit
solutions even at the bottom of the conformal window -- see
fig.~\ref{QCD_convergence_BZ_loops} here and fig.~7 in \cite{FP}.  
\end{description}

A crucial assumption we have made is that non-perturbative 
corrections are small within the conformal window so long as they 
are not implied by inconsistency of perturbation theory.
This assumption makes sense provided {\em no chiral symmetry breaking takes
place within the conformal window} (this is known to be the case in
SQCD). Usually when chiral symmetry is
broken, one expects quarks to develop dynamical masses. Then at
momentum scales below these masses quarks will decouple 
from the dynamics. This would clearly invalidate the
perturbative analysis that takes into account $N_f$ massless quarks.
In particular decoupling could make effective charges {\em non-monotonous} 
with scale and thus invalidate the notion of a single valued $\beta$ function.
In the absence of chiral symmetry breaking, and in particular within the 
conformal window, it is natural to expect that effective charges
evolve monotonously with scale. Thus the monotonicity property
conjectured for specific observables by various suggested
generalizations \cite{c-theorem-gen} of the c-theorem \cite{c-theorem}
could actually be a generic property
of all effective charges within the conformal window.
 
Consistency of perturbation theory together with the assumption
that also non-perturbative effects are small implies that QCD is, in
some sense, weakly coupled even at the bottom of the conformal window. 
The weak coupling nature of the QCD might be a disappointing message 
to anyone wishing to extend the notion of duality beyond the 
supersymmetric case.  On the other hand, the fact that the infrared coupling 
is basically controlled by the perturbative $\beta$ function opens up the
possibility to analyze various physical quantities in the infrared 
near the confining phase transition. An example of such analysis 
is that of \cite{Appelquist} which explains the phase transition at
the bottom of the window through decoupling of quarks due to
chiral symmetry breaking.

The analyticity structure of the coupling seems to be 
indicative of the reliability of perturbation theory in the
infrared in both QCD and SQCD. A priori one might suspect that not much
could be gained by analyzing the singularity structure of the coupling
which is scheme dependent.
It turns out, however, that several characteristics of the
$\beta$ function are generic. One simple example, which is crucial 
for our analysis, is the negative sign of $\beta_2$ for physical schemes (and
${\hbox{$\overline{\hbox{MS}}\,$}}$) in QCD in the entire 
conformal window. It is not clear to us how general this property is.

The most important universal infrared property of the $\beta$ function is
the critical exponent which turns out to be the key
parameter in the condition for a causal analyticity structure.
The condition $0\leq\gamma<1$ is necessary for a causal coupling 
at any loop order and even beyond perturbation theory, while in
practice at the 3-loop order it is usually also sufficient. 
$\gamma=0$ corresponds to a free theory, as obtained
at the top of the conformal window. In the supersymmetric case,
duality implies that $\gamma$ is the same in both theories, and thus
$\gamma$ vanishes also at the bottom of the window, where the magnetic theory
is free. 
In terms of the electric theory the vanishing of $\gamma$ can be
most simply interpreted as corresponding to a smooth change of
the $\beta$ function as $N_f/N_c$ crosses its critical value: the
$\beta(x)$ curves continuously change from curves that cross the
$x$ axis at two near-by points to ones that do not cross the $x$ axis
at all. The critical curve corresponding to
$\beta(x)$ at the bottom of the window just touches the $x$ axis from below: 
$\gamma=0$ implies a double zero.  

In QCD we trust perturbation theory down to the bottom of the window
and find that $\gamma\neq 0$ there.
On the other hand, it is clear that for some effective charges, 
such as the one associated with the potential between heavy quarks,
there is no scale invariant behavior at large distances in the
confining phase and hence the fixed-point must disappear as the lower 
boundary of the conformal window is crossed.
The disappearance of the fixed-point with a non-vanishing $\gamma$ 
implies that the above picture of a smooth change of the $\beta$
function cannot be realized. In this case, when the lower boundary of the
window is crossed there is a discontinuous change in the form of the
$\beta$ function for $x>x_{\FP}$. 
This discontinuity, however, does not imply a jump in the
coupling $x(Q^2)$ at any finite $Q^2$ but only at the infrared limit
itself.   
Finding that in QCD $\gamma$ can be perturbatively calculated and   
that it does not vanish at the bottom of the conformal window 
is a further indication of the absence of a dual description of the
infrared in terms of some other weakly coupled fields.

\vskip 30pt
 
\begin{flushleft}
{\large\bf Acknowledgments}
\end{flushleft}
The authors are grateful to D. Anselmi for very interesting and
helpful discussions.
E. G. thanks also A. Armoni, N. Itzhaki and  M. Karliner  
for very useful discussions.
This research was supported in part by the Israel
Science Foundation administered by the Israel Academy of Sciences and
Humanities, by a Grant from the G.I.F., the German-Israeli
Foundation for Scientific Research and Development, by the Charles
Clore doctoral fellowship and by the EC program `Training and Mobility
of Researchers', Network `QCD and Particle Structure', 
contract ERBFMRXCT980194.

\newpage
\begin{flushleft}
{\large\bf Appendix -- The analyticity structure of the NSVZ coupling}
\end{flushleft}

In Sec. 3 we analyze the singularity structure of the 2-loop coupling
in SQCD. Another natural choice for the SQCD $\beta$ function, at the same
level of approximation, is the NSVZ form
with the matter field anomalous dimension calculated to 
first order \cite{NSVZ,SV}. It is interesting to see whether the
condition for causality of the coupling with this choice for the
$\beta$ function agrees with that of the 2-loop choice.

The NSVZ $\beta$ function at this order is:
\beq
\beta(x)=-B_0x^2\,\frac{1-E_1x}{1-Dx}
\label{NSVZ_beta}
\eeq
where 
$B_0$ is given in (\ref{B0}), 
\beq
E_1=\frac{N_f}{3N_c-N_f}\,\frac{N_c^2-1}{2N_c}
\label{E}
\eeq
and 
\beq
D=N_c/2.
\label{D}
\eeq
The $\beta$ function in the NSVZ form yields, of course, the scheme
invariant 2-loop $\beta$ function of (\ref{beta_SQCD}) upon expansion,
with $C_1=D-E_1$ (see eq. (\ref{C1})).

\setcounter{footnote}{0}
Exact integration of the $\beta$ function (\ref{NSVZ_beta}) yields
a coupling constant that can be written explicitly as a function of the
scale parameter $t=\ln(Q^2/\Lambda^2)$ using the Lambert W
function\footnote{This solution is very similar to the one
presented in ref. \cite{LamW} for the 2-loop and Pad\'e improved 3-loop
coupling.}:
\beq
\begin{array}{c}
\displaystyle
x(Q^2)=\frac{1}{E_1}\,\,\frac{1}{1+G W(z)} \nonumber\\
\phantom{a}\\
\displaystyle
z = \frac{1}{G}\exp\left[\frac{1}{G}\left(-1\,+\, \frac{B_0}{E_1} t\right)\right]
\end{array}
 \label{W_sol_NSVZ}
\eeq
where $W(z)$ is defined by \cite{Lambert}:
\beq
W(z) \exp\left[W(z)\right]=z.
\label{LambertW_def}
\eeq
and 
\beq
G\equiv 1-\frac{D}{E_1}.
\eeq

Next, one should specify the branch of the Lambert W function such that 
asymptotic freedom will be obeyed ($x$ should be real and positive
at large (space-like) $Q^2>0$ and approach zero as
$t\longrightarrow\infty$). Since we are
interested in the asymptotically free case where $B_0>0$, we find that
also $E_1>0$ and thus the sign of $G$ determines both the sign 
of $z$ and its magnitude in the ultraviolet:
\begin{description}
\item{(a) }
if $G>0$, $z>0$ and approaches infinity in the ultraviolet
($z\longrightarrow 0$ in the infrared). Then the
relevant branch on the space-like axis is the principle branch 
$W_0(z)$, which is a monotonically increasing positive function of $z$
for $z>0$; $W_0(z)\longrightarrow\infty$ as $z\longrightarrow \infty$ 
thereby assuring asymptotic freedom.  $G>0$ is obtained whenever $E_1>D>0$.
The singularity structure in the $Q^2$ plane can be analyzed following
the lines of \cite{LamW}. There are two possibilities (for $G>0$): if
$B_0/(E_1G)>1$, i.e. if $B_0>E_1-D$, there is a pair of Landau branch
points at complex $Q^2$ values. On the other hand if $B_0/(E_1G)<1$,
i.e. $B_0<E_1-D$, there are no Landau singularities and the
analyticity structure is consistent with causality.
\item{(b) } if $G<0$, $z<0$ and approaches zero in the ultraviolet
  ($z\longrightarrow -\infty$ in the infrared). In this case, the
  relevant branch is $W_{-1}(z)$ which turns to minus infinity for
  $z\longrightarrow 0^-$, thereby assuring asymptotic freedom. The
  singularity structure is simple: there is a single Landau branch
  point at $z=-1/e$ corresponding to a certain $Q^2>0$ on the
  space-like axis ($t_{sing}=(E_1/B_0)[1-G+G\ln(-G)]$).
\end{description}
Note that contrary to the 2-loop or Pad\'e improved 3-loop couplings
in QCD, a simple pole from the denominator of (\ref{W_sol_NSVZ}) does
not appear, since whenever $W$ is real, its sign is the same as that
of $G$, and thus it is guaranteed that $1+GW>0$.

We conclude that the condition for a causal coupling is $G>0$ and
$B_0/(E_1G)<1$, where in fact the more restrictive condition is the
second one. Thus, the NSVZ coupling is causal whenever 
\beq
B_0<E_1-D.
\label{causal_NSVZ}
\eeq
The condition (\ref{causal_NSVZ}) {\em coincides} with the condition for
causality of the 2-loop coupling $B_0<-C_1$. Similarly, the condition
to have a space-like Landau singularity in the NSVZ coupling, $D>E_1$,
coincides with the corresponding condition in the 2-loop coupling: $C_1>0$.

\newpage

\newpage

\newpage

\newpage

\newpage

\newpage

\newpage

\newpage

\newpage

\newpage

\newpage


\begin{thebibliography}{99}

\bibitem{BZ}  T. Banks and A. Zaks, {\it Nucl. Phys.} {\bf B196}
  (1982) 189.

\bibitem{oneloop}  D. Gross and F. Wilczek, {\it Phys. Rev. Lett.}
{\bf 30} (1973) 1343;  H.D. Politzer {\it Phys. Rev. Lett.} {\bf 30}
(1973) 1346.

\bibitem{twoloops}  W. E . Caswell, {\it  Phys. Rev. Lett.} {\bf
33} (1974) 244; D. R. T. Jones, {\it Nucl. Phys. }{\bf B75} (1974) 531.

\bibitem{Oehme_metric_confinement}
R. Oehme, W. Zimmermann {\em Phys. Rev.} {\bf D21} (1980) 471;
{\em Phys. Rev.} {\bf D21} (1980) 1661;
R. Oehme, {\em Phys. Lett.} {\bf 195B} (1987) 60; 
{\em Phys. Rev.} {\bf D42} (1990) 4209-4221;
{\em Phys. Lett.} {\bf B252} (1990) 641-646;
K. Nishijima {\em Prog. Theor. Phys.} {\bf 75} (1986) 1221.

\bibitem{Oehme_potential_confinement}
K. Nishijima, {\it Prog. Theor. Phys.} 77 (1987) 1035;
R. Oehme, {\it Phys. Lett.} {\bf B232} (1989) 498.

\bibitem{Appelquist} T. Appelquist, J. Terning and L.C.R. Wijewardhana
{\it Phys. Rev. Lett.} {\bf 77} (1996) 1214-1217;  
T. Appelquist, A.Ratnaweera, J. Terning and L.C.R. Wijewardhana,
{\it The phase structure of an SU(N) gauge theory with $N_f$ flavors},
YCTP-P15-98, hep-ph/9806472.

\bibitem{Shuryak}
M. Velkovsky, E. Shuryak {\em QCD with large number of quarks: effects
  of the instanton - anti-instanton pairs}, SUNY-NTG-96-37, hep-ph/9703345.

\bibitem{Miransky} V.A. Miransky and Koichi Yamawaki,
{\it Phys. Rev.} {\bf D55} (1997) 5051-5066 (Erratum-ibid {\bf D56} 
(1997) 3768).

\bibitem{lattice_old} J.B. Kogut and D.R. Sinclair, 
{\it Nucl. Phys.} {\em B295} [FS 21] (1988) 465;
F.R. Brown, H.Chen, N.H. Christ, Z. Dong, R.D. Mawhinney, W.Shafer and 
A. Vaccarino, {\it Phys. Rev.} {\bf D46} (1992) 5655.

\bibitem{lattice_Japan} Y. Iwasaki, {\em Phase structure of lattice QCD for
    general number of flavours} hep-lat/9707019 and references
  therein.

\bibitem{FP} E. Gardi and M. Karliner, 
{\it Nucl. Phys.} {\bf B529} (1,2) (1998) 383-423.

\bibitem{LamW} E. Gardi, G. Grunberg and M. Karliner,
JHEP {\bf 07} (1998) 007.

\bibitem{ECH}
G. Grunberg, {\it Phys. Lett.} {\bf 95B} (1980) 70, 
Erratum-ibid.{\bf 110B} (1982) 501; 
{\it Phys. Rev.} {\bf D29} (1984) 2315.

\bibitem{BZ_grunberg} G. Grunberg, {\it Phys. Rev.} {\bf D46} (1992)
  2228.

\bibitem{CaSt}  S. A. Caveny and P. M. Stevenson, {\em The Banks-Zaks
Expansion and ``Freezing'' in Perturbative QCD}, hep-ph/9705319.

\bibitem{Seiberg} N. Seiberg, {\it Nucl. Phys.} {\bf B435} (1995) 129.

\bibitem{Peskin} M.E. Peskin {\em Duality in Supersymmetric 
Yang-Mills Theory}, SLAC-PUB-7393, hep-th/9702094.

\bibitem{Shifman_review}  M. Shifman, {\it Prog. Part. Nucl. Phys.} 
{\bf 39} (1997) 1-116.

\bibitem{Oehme_SQCD} R. Oehme, {\it Phys. Lett.} {\bf B399}
(1997) 67.

\bibitem{superconvergence_SQCD}
M. Tachibana, {\it Phys. Rev.} {\bf D58} (1998) 045015.

\bibitem{Baker}
George A. Baker, Jr. {\em Essentials of Pad\'e Approximants }.
Academic Press, inc. (London), 1975. See Chapter 8 for a discussion of
N-point Pad\'e Approximants.

\bibitem{Lambert}
R.M.~Corless, G.H.~Gonnet, D.E.G.~Hare,
  D.J.~Jeffrey and D.E.~Knuth, ``On the Lambert W function'',
{\em Advances in Computational Mathematics}, {\bf 5} (1996) 329,
\hbox{
available from
{\tt
http://pineapple.apmaths.uwo.ca/{\mysim}rmc/papers/LambertW/}.}

\bibitem{private}
N.G. Uraltsev, private communication with G. Grunberg.

\bibitem{threeloops}  O. V. Tarasov, A. A. Vladimirov, and A.
Yu. Zharkov, {\it Phys. Lett.} {\bf B93} (1980) 429.

\bibitem{fourloops}  J. A. M. Vermaseren, S. Larin and T. van Ritbergen,
{\it Phys. Lett.} {\bf B400} (1997) 379-384.

\bibitem{V-scheme-corrected} Y. Sch\"oder, 
{\it The Static Potential in QCD},
DESY 98-191, hep-ph/9812205.

\bibitem{DRED} I. Jack, D.R.T Jones and A. Pickering,
{\it The connection between the DRED and NSVZ Renormalization Schemes},
to be published in {\it Phys. Lett.} {\bf B}, LTH 426, hep-ph/9805482.

\bibitem{NSVZ} V.A. Novikov, M.A. Shifman, A.I. Vainshtein and V.I. Zakharov,
{\it Nucl. Phys.} {\bf B229} (1983) 381.

\bibitem{SV} M.A. Shifman and A.I. Vainshtein, {\it Nucl. Phys.} {\bf B277} 
(1986) 456; {\it Nucl. Phys.} {\bf B359} (1991) 571-580. 

\bibitem{KSV} I.I. Kogan, M. Shifman and A. Vainshtein
{\it Phys. Rev.} {\bf D53} (1996) 4526-4537.

\bibitem{Oehme_reduction} R. Oehme, {\it Reduction of Dual Theories},
hep-th/9808054.

\bibitem{Gross} D. Gross, {\it Methods in Field Theory}, Les Houches
1975, R. Ballian and J. Zinn-Justin editors, North-Holland Pub. (1976).   

\bibitem{Chyla} Jir\'i Ch\'yla, {Phys. Rev.}{\bf D38} (1988) 3845.

\bibitem{ee_duality} A. de Gouv\^ea, A. Friedland, and
H. Murayama, {\it Seiberg Duality and $e^+e^-$ Experiments},
hep-th/9810020.

\bibitem{Anselmi} D. Anselmi, M. Grisaru and  A. Johansen,
{\it Nucl. Phys.} {\bf B491} (1997) 221-248.

\bibitem{preparation} E. Gardi and G. Grunberg, {\em Extension 
of Seiberg Duality to the vicinity of the infrared fixed-point}, 
in preparation.

\bibitem{Why}
E. Gardi,
{\em Phys. Rev.} {\bf D56} (1997) 68.

\bibitem{c-theorem} A.B. Zamalodchikov, JETP Lett. 43 (1986) 730.

\bibitem{c-theorem-gen} J.L. Cardy, {\it Phys. Lett.} {\bf B 215} (1988)
749; D. Anselmi, ``Anomalies, Unitarity and Quantum Irreversibiliy'',
CERN-TH/99-33, hep-th/9903059 and refs. therein.

\end{thebibliography}
\end{document}